\documentclass[10pt,superscriptaddress,amsmath,amssymb,aps,prx,twocolumn,a4paper,floatfix,longbibliography]{revtex4-1}

\usepackage[version=3]{mhchem}
\usepackage{graphicx,times,bm,amssymb,amsmath,booktabs,natbib,physics,subfigure,makecell,hyperref,multirow}

\def\be{\begin{equation}}       \def\ee{\end{equation}}
\def\bea{\begin{eqnarray}}      \def\eea{\end{eqnarray}}

\def\degree{${}^{\circ}$}
\begin{document}
\title{Ni-based transition-metal trichalcogenide monolayer: a strongly correlated quadruple-layer graphene}

\author{Yuhao Gu}
\affiliation{Beijing National Laboratory for Molecular Sciences, State Key Laboratory of Rare Earth Materials Chemistry and Applications, Institute of Theoretical and Computational Chemistry, College of Chemistry and Molecular Engineering, Peking University, 100871 Beijing, China}
\affiliation{Beijing National Laboratory for Condensed Matter Physics,
and Institute of Physics, Chinese Academy of Sciences, Beijing 100190, China}

\author{Qiang Zhang}
\affiliation{Beijing National Laboratory for Condensed Matter Physics,
and Institute of Physics, Chinese Academy of Sciences, Beijing 100190, China}

\author{Congcong Le}
\affiliation{Kavli Institute of Theoretical Sciences, University of Chinese Academy of Sciences,
Beijing, 100190, China}
\affiliation{Beijing National Laboratory for Condensed Matter Physics,
and Institute of Physics, Chinese Academy of Sciences, Beijing 100190, China}

\author{Yinxiang Li}
\affiliation{Beijing National Laboratory for Condensed Matter Physics,
and Institute of Physics, Chinese Academy of Sciences, Beijing 100190, China}

\author{Tao Xiang} 
\affiliation{Beijing National Laboratory for Condensed Matter Physics,
and Institute of Physics, Chinese Academy of Sciences, Beijing 100190, China}
\affiliation{Kavli Institute of Theoretical Sciences, University of Chinese Academy of Sciences,
Beijing, 100190, China}
\affiliation{Collaborative Innovation Center of Quantum Matter,
Beijing, China}

\author{Jiangping Hu}\email{jphu@iphy.ac.cn}
\affiliation{Beijing National Laboratory for Condensed Matter Physics,
and Institute of Physics, Chinese Academy of Sciences, Beijing 100190, China}
\affiliation{Kavli Institute of Theoretical Sciences, University of Chinese Academy of Sciences,
Beijing, 100190, China}
\affiliation{Collaborative Innovation Center of Quantum Matter,
Beijing, China}

\begin{abstract}
We investigate the electronic physics of layered Ni-based  trichalcogenide NiPX$_3$ (X=S, Se), a member of  transition-metal trichalcogenides (TMTs) with the chemical formula, ABX$_3$.  These Ni-based TMTs distinguish themselves from other TMTs as their low energy electronic physics can be effectively described by the two $e_g$ d-orbitals.  The major band kinematics is characterized by the unusal long-range effective hopping between  two third nearest-neighbor (TNN) Ni sites in the two-dimensional Ni honeycomb lattice so that the Ni lattice can be equivalently viewed as four weakly coupled   honeycomb sublattices. Within each sublattice, the electronic physics is described by a strongly correlated two-orbital graphene-type model that results in an antiferromagnetic (AFM) ground state near half filling.  We show that the low energy physics in a paramagnetic state is determined by the eight Dirac cones which locate at $K$, $K'$, $\frac{K}{2}$ and $\frac{K'}{2}$ points in the first Brillouin zone with a strong AFM fluctuation between two $K (K')$ and $\frac{K'}{2} (\frac{K}{2})$ Dirac cones and carrier doping can sufficiently suppress the long-range AFM order and allow other competing orders, such as superconductivity,  to emerge. The material can be an ideal system to study many exotic phenomena emerged from strong electron-electron correlation, including a potential $d\pm id$ superconducting state at high temperature.
\end{abstract}


\maketitle

\section{Introduction}
Since the discovery of graphene\cite{graphene2004} a decade ago, two-dimensional (2D) materials have been a research frontier for both fundamental physics and practical device applications\cite{wang_new_2018,susner_metal_2017}.
 Transition-metal trichalcogenides (TMTs) with the chemical formula ABX$_3$ (X=S, Se, Te), which  were known more than a century ago\cite{friedel1894thiohypophosphates,ferrand1895bull},  are layered van der Walls (vdW) materials.  Recently,  this family of materials  has attracted great research attention as potential excellent
candidates  to explore 2D magnetism  for novel spintronics applications.

All the members in the family of \ce{ABX3} materials are built on a
 common structural unit,  \ce{(P2X6)^{4-}} (X=S, Se, Te) anion complex.  However, the cation atom $A$ is rather flexible,  ranging from vanadium to zinc (A=V, Cr, Mn, Fe, Co, Ni, Zn, etc.) in the row of the 3$d$ transition metal,   partial alkaline metal in group-II, and some other metal ions. As shown in Fig.\ref{lattice} (a,b), the cation is coordinated with six chalcogen anions to form an octehedra complex.  In the two-dimensional layer, the cation forms a graphene-type honeycomb lattice. The transition-metal trichalcogenides exhibit a variety of intriguing magnetically ordered  insulating states\cite{le_flem_magnetic_1982}. Recently, under high pressure,  \ce{FePSe3} can also become a superconductor\cite{wang_emergent_2018}.

Among this family of materials, the Ni-based trichalcogenides can carry intriguing electronic physics, such as strong charge-spin coupling\cite{So-prl18}, because of the following reasons. First,  as the transition metal cation and chalcogen anions form an octahedral complex, the 3$d$-orbitals of the transition metal are divided into high energy $e_g$ and low energy $t_{2g}$ groups. In the case for Ni which has 8 electrons in the 3$d$-shell,  the $t_{2g}$ orbitals are fully occupied and the two $e_g$ orbitals are half-filled as shown in Fig.\ref{lattice} (c). The $t_{2g}$ orbitals are inactive. The Ni-based trichalcogenides should be described by  a relatively simpler low energy effective model than other materials. Second, unlike a two dimensional square lattice,   a honeycomb lattice easily exhibits a Dirac-cone type of energy dispersion. Near half filling, both one-orbital model, such as graphene\cite{graphene2004}, and two-orbital models\cite{wu_flat_2007,wu_p_2008} in the honeycomb lattice are featured with Dirac points near Fermi energy. With the strong electron-electron correlation in the 3$d$-orbitals, the Ni-based trichalcogenide thus can be a candidate of strongly correlated Dirac electron systems.  It is worth to mention that a recent major research effort has aimed to increase the electron-electron correlation in graphene\cite{cao2018correlated}, in which  flat bands have to be engineered to observe correlation effects because of  $p$-orbitals.  Finally,  both density functional theory (DFT) calculation and experimental measurements have suggested that the Ni honeycomb lattice forms  the zigzag antiferromagnetic insulating ground state featured as double parallel ferromagnetic chains being anti-ferromagnetically (AFM) coupled\cite{le_flem_magnetic_1982,chittari_electronic_2016}. The material offers a promising platform to study  the interplay between the low energy Dirac electronic physics and the magnetism. Such an interplay is believed to be responsible for many important phenomena, for example,  high temperature superconductivity in both cuprates and iron-based superconductors\cite{hu_local_2012,hu_identifying_2016}.

In this paper, we show that  the Ni-based TMTs are Dirac materials with strong electron-electron correlation.  Their low energy electronic physics can be entirely attributed to the two $e_g$ $d$-orbitals  with  a band kinematics dominated by an   unusual "long-range" hoppings between  two third nearest-neighbor (TNN) Ni sites in the  Ni honeycomb lattice. Thus, the original Ni lattice can be divided into four weakly coupled honeycomb sublattices. Within each sublattice, the electronic physics is described by a strongly correlated two-orbital graphene-type model. The couplings between four sublattices, namely, the nearest-neighbor (NN) and the second NN hoppings (SNN) in the original lattice, can be adjusted by applying external pressure or chemical methods. 
In the absence of the strong electron-electron correlation, the low energy physics  is determined by the eight Dirac cones which locate at $K(k^\prime)$ and three non-equivalent pairs of $\frac{K}{2}(\frac{K^\prime}{2})$ points in the  first Brillouin zone. In the presence of  strong electron-electron correlation, strong AFM  interactions arise  between two NN sites within each honeycomb sublattice. Namely, in the view of the original honeycomb lattice, the strong AFM interactions only exist between two TNN sites. Near half filling, Dirac cones are gapped out by the long-range AFM order. Using the standard slave-boson approach, we show that the doping can sufficiently suppress the long range AFM order. In a wide range of doping, a strong AFM fluctuation can exist between the Dirac cones and  a $d\pm id$ superconducting state can be developed.

 It is interesting to make an analogy between the above results and those known in high temperature superconductors, cuprates and iron-based superconductors.   For the latter, it is known that the dominant AFM interactions are between two NN sites in curpates and between two SNN sites in iron-based superconductors, which are believed to be responsible for the $d$-wave and the extended $s$-wave pairing superconducting states respectively\cite{Seo2008,hu_local_2012,hu_identifying_2016}. All these AFM interactions are generated through superexchange mechanism. Thus, the Ni-based TMTs, having the AFM interactions between two TNN sites, can potentially provide the ultimate piece of evidence to settle superconducting mechanism in unconventional high temperature superconductors. 

This paper is organized as follows. In Section \ref{section1}, we briefly review transition-metal phosphorous trichalcogenides  and specify our DFT computation methods for magnetism and band structures. In Section \ref{s2}, we analyze the band structure of the paramagnetic states, derive the tight-binding Hamiltonian and discuss the low energy physics near  Fermi surfaces.  In Section \ref{s3}, we calculate the magnetic states of the materials and derive the effective magnetic exchange coupling parameters.  In Section \ref{s4}, we use the slave-boson meanfield method to derive the phase diagram upon doping and discuss the possible superconducting states. In the last section, we make summary and discussion.



\section{transition-metal phosphorous trichalcogenides}
\label{section1}
\begin{figure}
\centerline{\includegraphics[width=0.5\textwidth]{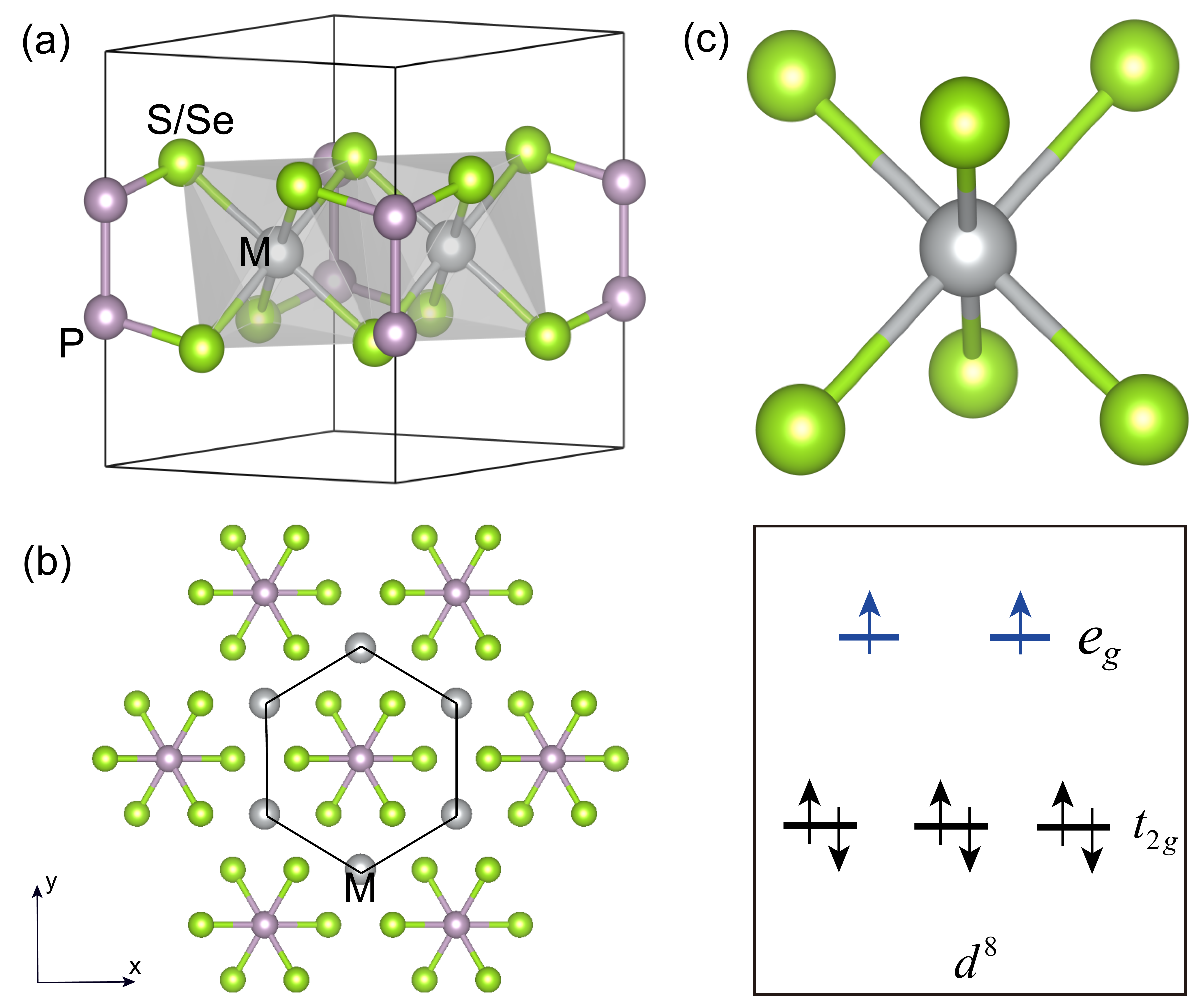}} \caption{(a) The crystal structures of the monolayer \ce{NiPX3} (X=S, Se) (space group P-31m). (b) The top view  of the monolayer \ce{NiPX3}. (c) The octahedral crystal field splitting of Ni atoms.
 \label{lattice} }
\end{figure}

The \ce{MPX3} metal phosphorous trichalcogenides (M=Mg, Sn, Sc, Mn, Fe, Co, Ni, Cd, etc and X=S, Se) are a famous family of 2D van der Waals (vdW) materials\cite{wang_new_2018,susner_metal_2017}. The bulk \ce{MPX3} crystal consists of AA-stacked or ABC-stacked single-layer assemblies which are held together by the vdW interaction. The vdW gap distance of the \ce{MPX3} with 3$d$ transition metal elements is about 3.2 \AA, much wider than the well studied \ce{MoS2}-type 2D vdW materials\cite{Mos2}, which indicates that the vdW interaction in \ce{MPX3} is relatively weak. The monolayer structure of \ce{MPX3} is constructed by \ce{MX6} edge shared octahedral complexes. Similar to the \ce{MoS2}-type \ce{MX2} materials, the monolayer \ce{MPX3} can be considered as the monolayer \ce{MX2} with one third of M sites substituted by \ce{P2} dimers, i.e., \ce{MPX3} can be considered as \ce{M_{2/3} (P2)_{1/3}X2}. Thus, the triangle lattice in \ce{MX2} transforms to the honeycomb lattice in \ce{MPX3} with the \ce{P2X6^{4-}} anions being located at the center of the honeycomb. The P-P and P-X bond lengths indicate that the P-P and P-X bonds are covalent bonds in \ce{P2X6^{2-}} anions.  As the transition-metal atoms are in octahedral environment, the five $d$ orbitals are split into two groups, $e_g$ and $t_{2g}$, as shown FIG.\ref{lattice} (c). 

With the weak vdW interaction, the essential electronic physics in \ce{MPS3} is determined within a monolayer. In fact, the atomically thin \ce{FePS3} layer has been experimentally synthesized\cite{lee_ising-type_2016}. In this paper, without further specification, our study and calculations are performed for the monolayer as shown in Fig.\ref{lattice} (a) and (b). We use the monolayer structures cleaved from the experimental crystal structures in ICSD\cite{icsd_2007}. The experimental structural parameters of the \ce{MPX3}  are listed in the Table.\ref{str-table}.

\begin{table}
\caption{\label{str-table}%
The structural parameters for monolayer \ce{MPX3} (space group $P-31m$). }
\begin{ruledtabular}
\begin{tabular}{cccccc}
System  & a (\AA) & M-X (\AA) & P-X (\AA) & P-P (\AA) & M-X-M ($^{\circ}$)\\
 \colrule
 \ce{MnPS3}\cite{ouvrard1985structural} & 6.08 & 2.63 & 2.03 & 2.19 & 83.9 \\
 \ce{FePS3}\cite{ouvrard1985structural} & 5.94 & 2.55 & 2.02 & 2.19 & 84.7 \\
 \ce{CoPS3}\cite{klingen1968hexathio} & 5.91 & 2.51 & 2.04 & 2.17 & 85.5 \\
 \ce{NiPS3}\cite{rao1992magnetic} & 5.82 & 2.50 & 1.98 & 2.17 & 84.4 \\
 \ce{NiPSe3}\cite{brec1980proprietes}& 6.13 & 2.61 & 2.09 & 2.24 & 85.2 
\end{tabular}
\end{ruledtabular}
\end{table}

Our DFT calculation is  performed for the monolayer \ce{MPX3} structures together with built-in 20 \AA$ $ thick vacuum layers. 
We employ the Vienna ab initio simulation package (VASP) code\cite{kresse1996} with the projector augmented wave (PAW) method\cite{Joubert1999FromMethod} to perform the DFT calculations. The Perdew-Burke-Ernzerhof (PBE)\cite{perdew_generalized_1996} exchange-correlation functional was used in our calculations. Through out this work, the kinetic energy cutoff (ENCUT) is set to be 500 eV for the expanding the wave functions into a plane-wave basis and the $\Gamma$-centered k-mesh is $16\times16\times1$ for the nonmagnetic unit cell. The energy convergence criterion is $10^{-6}$ eV. We perform the static self-consistent calculation with the monolayer structure cleaved from the experimental crystal structures in ICSD\cite{icsd_2007}. Since there is no experimental data about \ce{CuPS3} with this layered hexagonal structure,  we  optimize \ce{CuPS3} with the force convergence criterion of 0.01 eV/\AA. In the study of effective Hamiltonian, we employ Wannier90\cite{mostofi2008wannier90} to calculate the hopping parameters of the tight binding model.
In the study of magnetism of \ce{MPX3}, the GGA plus on-site repulsion $U$ method (GGA+$U$) in the formulation of Dudarev \textit{et al.}\cite{Dudarev1998Electron-energy-lossStudy} is employed to describe the associated
electron-electron correlation effect. The effective Hubbard $U$ ($U_{eff}$) is defined by $U_{eff}=U-J_{Hund}$. In order to describe different magnetic orders, we 
build $2\times1\times1$ supercell and the k-mesh is $8\times16\times1$, correspondingly.

\section{Electronic band structures and  the tight binding model for  Ni-trichalcogenides }\label{sec:ele}
\label{s2}
\begin{figure}
\centerline{\includegraphics[width=0.5\textwidth]{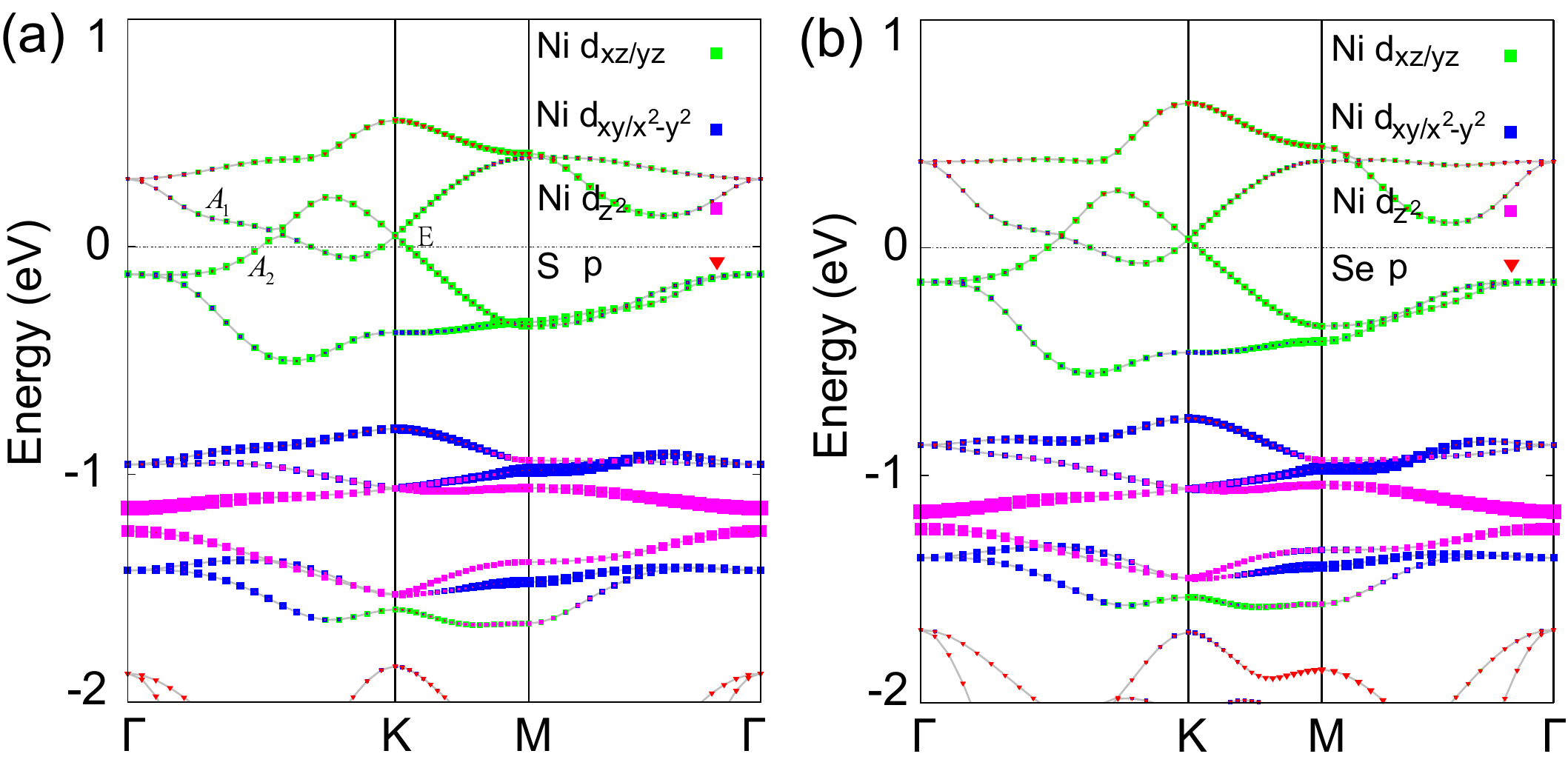}} \caption{Electronic band structures of (a) \ce{NiPS3} and (b) \ce{NiPSe3}.  The orbital characters of bands are represented by different colors. 
 \label{nm_band} }
\end{figure}

In Fig.\ref{nm_band}, we plot the band structures of \ce{NiPS3} and \ce{NiPSe3}, which are very similar to each other.    From Fig.\ref{nm_band}, it is clear that the five $d$-orbital bands are divided into two groups separated by a large crystal field splitting energy. The groups at the high and low energy are attributed to the two $e_g$ and three $t_{2g}$ orbitals respectively. The bands from  $t_{2g}$ orbitals are completely filled while the four bands from the two $e_g$ orbitals are close to half-filling.  This is consistent with the fact that  the \ce{Ni^{2+}} cations are six-coordinated with an octahedral geometry and the $d^8$ configuration in \ce{Ni^{2+}} contributes two electrons to the two $e_g$ orbitals. The physics near Fermi energy is controlled by the two $e_g$ $d_{xz/yz}$ orbitals. It is also worthy noting that the contribution of the S/Se-$p$ orbitals is considerable, especially near $\Gamma$ point, which indicates strong $d$-$p$ hybridization.

 Similar to the single orbital band structure in graphene, the two $e_g$ orbital bands are featured by Dirac points as well. Here, there are eight Dirac points which locate  at  $K$ and $K^\prime$ points, as well as around $K/2$ and $K^\prime/2$ points along $\Gamma$-$K$ line as shown in Fig.\ref{nm_band}.  These Dirac points are rather robust. To show the robustness, we analyze the symmetric character of the bands, namely, their irreducible representations. The two crossing bands along $\Gamma$-$K$ line belong to A$_1$ and A$_2$ irreducible representations, and the bands at $K$ points belong to E irreducible representation. Thus, the  Dirac points are protected by the symmetry as the bands belong to different representations.
 
\begin{table}
\caption{\label{hops}%
The NN, SNN and TNN hopping parameters and the band widths for \ce{NiPS3} and \ce{NiPSe3}; xz and yz represent the band indexes of the hopping parameters. }
\begin{ruledtabular}
\begin{tabular}{ccc}
  & \ce{NiPS3}& \ce{NiPSe3}\\
 \colrule
 $t^{NN}_{xzxz}$ (eV) & -0.050971 & -0.059617\\
 $t^{NN}_{yzyz}$ (eV) & -0.036294 & -0.014879\\
 $t^{SNN}_{xzxz}$ (eV) & 0.012118 & 0.019696\\
 $t^{SNN}_{yzyz}$ (eV) & -0.015141 & -0.017257\\
 $t^{SNN}_{xzyz}$ (eV) & 0.003175 & 0.003827\\
 $t^{TNN}_{xzxz}$ (eV) & -0.020218 & -0.019138\\
 $t^{TNN}_{yzyz}$ (eV) & 0.238574 & 0.256818\\
 band widths $e_g$ & 0.89 & 1.01
\end{tabular}
\end{ruledtabular}
\end{table}

\begin{figure}
\centerline{\includegraphics[width=0.5\textwidth]{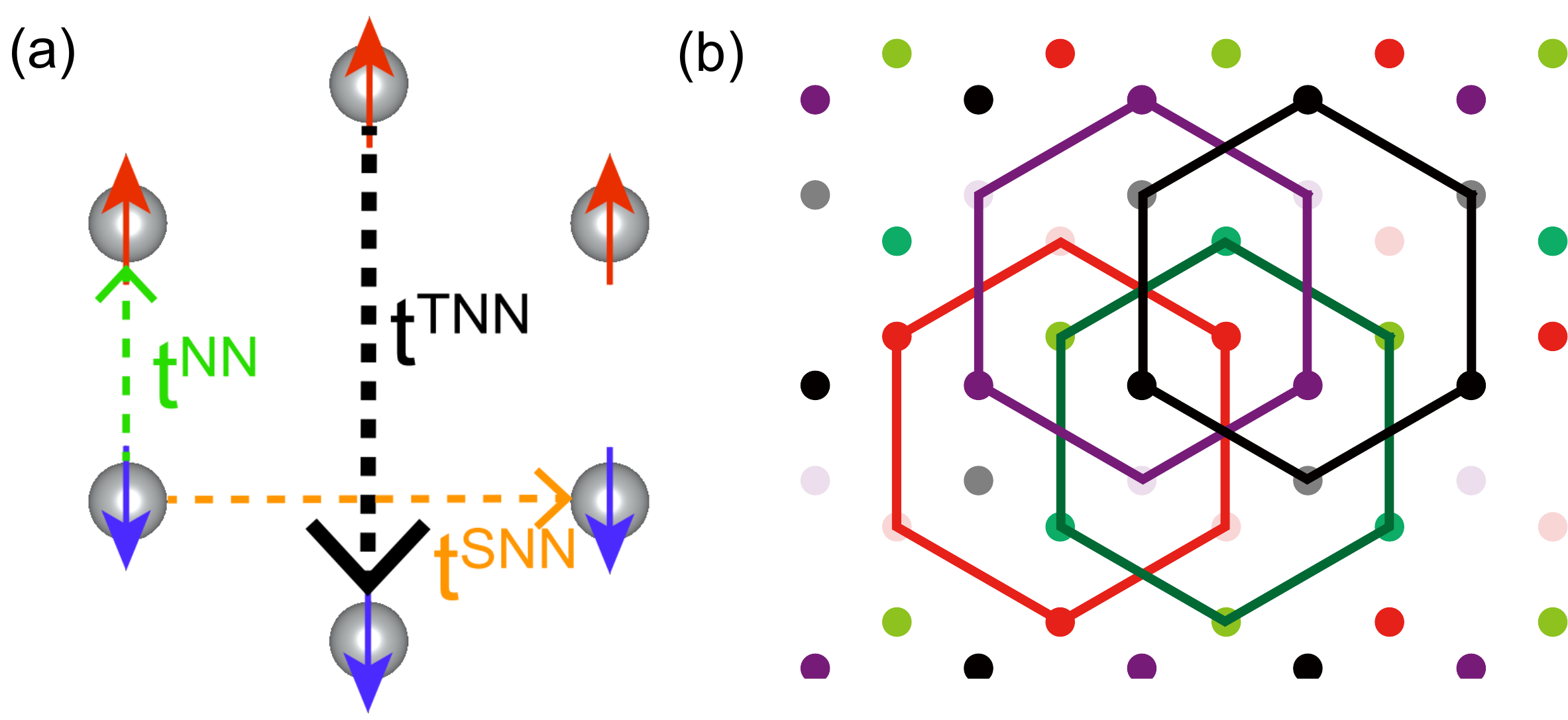}} \caption{(a) The three NN, SNN, and TNN hopping parameters marked by the green, brown and black dashed arrows, respectively and the zigzag AFM order with onsite red/blue arrows indicating spin up/down; (b) The four honeycomb sublattices.}
\label{sublattice}
\end{figure}

\begin{table}
\caption{\label{hops_angles}%
The NN, SNN and TNN hopping parameters for \ce{NiPS3} with different Ni-S-Ni angles; xz and yz represent the band indexes of the hopping parameters. }
\begin{ruledtabular}
\begin{tabular}{cccc}
Ni-S-Ni ($^{\circ}$) & 84.4 & 81.9 & 87.1 \\
 \colrule
 $t^{NN}_{xzxz}$ (eV) & -0.050971 & -0.038045 & -0.066167 \\
 $t^{NN}_{yzyz}$ (eV) & -0.036294 & -0.026956 & -0.040663 \\
 $t^{SNN}_{xzxz}$ (eV) & 0.012118 & 0.011367 & 0.013838 \\
 $t^{SNN}_{yzyz}$ (eV) & -0.015141 & -0.009442 & -0.021795 \\
 $t^{SNN}_{xzyz}$ (eV) & 0.003175 & 0.001937 & 0.004137 \\
 $t^{TNN}_{xzxz}$ (eV) & -0.020218 & -0.021682 & -0.019388 \\
 $t^{TNN}_{yzyz}$ (eV) & 0.238574 & 0.210725 & 0.269932
\end{tabular}
\end{ruledtabular}
\end{table}
In order to capture the two-dimensional electronic physics near the Fermi level, we construct the tight binding  Hamiltonian based on the two $e_g$ orbitals. The Hamiltonian can be written as 
\begin{equation}
    H_0=\sum_k \psi^\dagger_k h_k \psi_k,
    \label{h0}
\end{equation}
where the basis $\psi^\dagger_k= (a^\dagger_{xk},a^\dagger_{yk},b^\dagger_{xk},b^\dagger_{yk})$ and 
\begin{align}
    h_k=\left (\begin{array}{cc}
         \omega_k-\mu & \gamma_k \\
        \gamma_k^\dagger & \omega_k^T -\mu
    \end{array}\right)\label{hk}
\end{align}
with $\mu $ being the chemical potential and
\begin{align}
\omega_k&=\sum_je^{-i\mathbf{k}\cdot  \mathbf{a_{1j}}} T_j^{SNN}\\
\gamma_k&=\sum_{j}e^{-i\mathbf{k}\cdot \mathbf{a_{2j}}}T_j^{NN}+e^{-i\mathbf{k}\cdot \mathbf{a_{3j}}}T_j^{SNN}.
\end{align}
Here $a^\dagger_{xk} (b^\dagger_{yk})$ is the electron annihilator operator of orbital $xz (yz)$ in  the usual $A (B)$ sublattice of the honeycomb lattice and vectors $\mathbf{a_1,a_2,a_3}$ are the first, second and third neighbor vectors.   $T_j^i=C_{3j}T^iC^{-1}_{3j}$ is the $i$-th neighbor hopping matrix via the bond along ij bond direction and $C_{3j}$ is the threefold rotation operation to the ij direction relative to the initial setting.  $T^i(i=NN,SNN,TNN)$ is the hopping matrix with the direction marked in Fig.\ref{sublattice} (a):
\begin{align}
T^{i}=\left (\begin{array}{cc}
t^{i}_{xzxz} &t^{i}_{xzyz}\\-t^{i\ast}_{xzyz} & t^{i}_{yzyz}
\end{array}\right).
\end{align}
By the lattice symmetry, $t^{NN}_{xzyz}=t^{TNN}_{xzyz}=0$.  We will use eV as the energy unit for all hopping parameters. The results of \ce{NiPS3} and \ce{NiPSe3} are similar, as shown in Table.\ref{hops}.  The explicit formula of the Hamiltonian is given in  Appendix A. Here we focus on the results of \ce{NiPS3}.

\begin{figure}
\centerline{\includegraphics[width=0.5\textwidth]{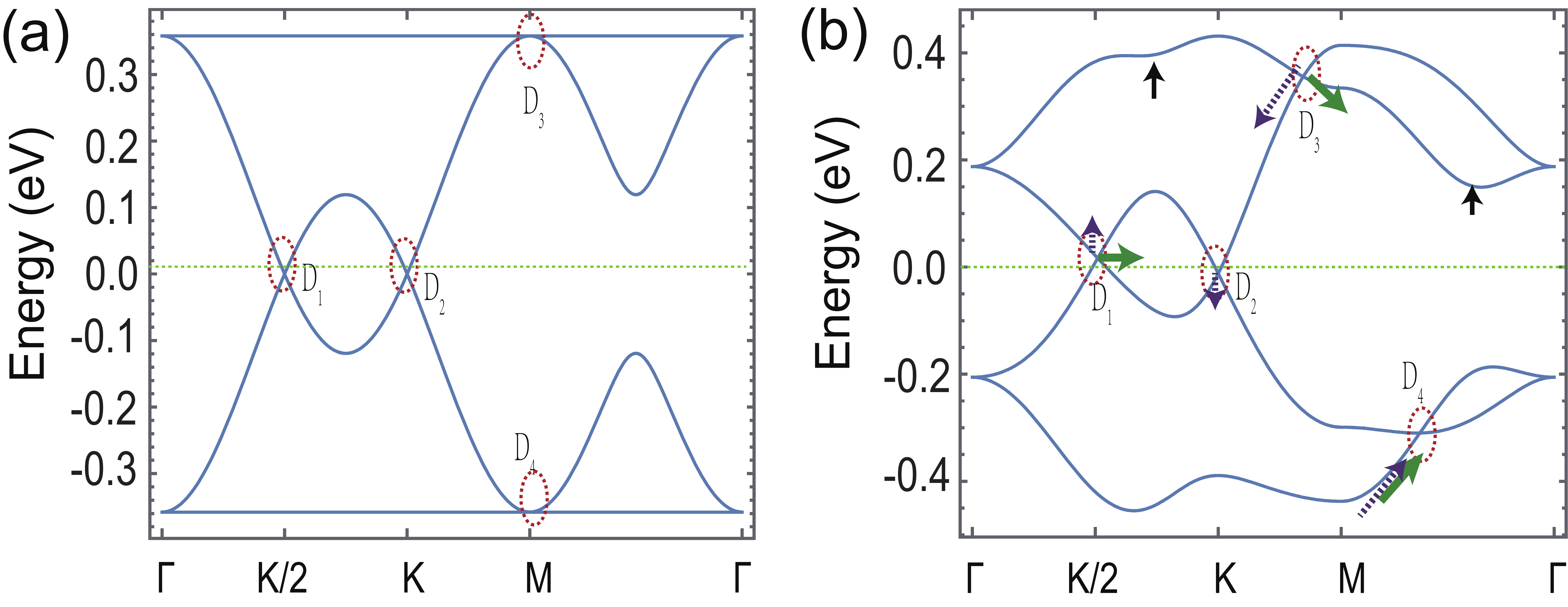}} \caption{The effect of different hopping parameters on the band structure and Dirac points. (a) The band dispersion with only the leading TNN hopping  $t^{TNN}_{xzxz}$; (b) The band dispersion with all hopping parameters: the green, purple and black arrows represent the motion of the band and Dirac points by increasing the NN, SNN and TNN hopping parameters, respectively.}
\label{hopping-tnn}
\end{figure}

It is  interesting to notice that the leading term in the above Hamiltonian is $t^{TNN}_{yzyz}$, the TNN $\sigma $-bond hoppings as shown in Fig.\ref{sublattice} (a), which is almost one order of magnitude larger than the other hopping parameters, namely, the NN and SNN hopping parameters. Thus, we can consider these TNN hoppings as the dominant hopping parameters and treat other hoppings as perturbations. In Fig.\ref{hopping-tnn} (a), we plot the band dispersion with only the TNN hopping parameters.  With only these TNN hoppings, the original Ni honeycomb lattice is divided into four decoupled sublattices as shown in Fig.\ref{sublattice} (b). Within each honeycomb sublattice, the model is identical to the one previously studied in an ultracold atomic honeycomb lattice with two degenerate p-orbitals\cite{wu_flat_2007,wu_p_2008}. As shown in Fig.\ref{hopping-tnn} (a), there are two completely flat bands and two dispersive bands. The flat bands stem from the localized binding and anti-binding molecular orbitals\cite{wu_flat_2007}. The two dispersive bands create the eight Dirac points. With only these TNN hoppings, the second pair of Dirac points are exactly located at $K/2$ and $K^\prime/2$ points. This pair is simply created through the Brillouin zone folding because of the sublattice structure. Thus, the presence of the two pairs of Dirac points underlines the sublattice structure. 

\begin{figure}
\centerline{\includegraphics[width=0.5\textwidth]{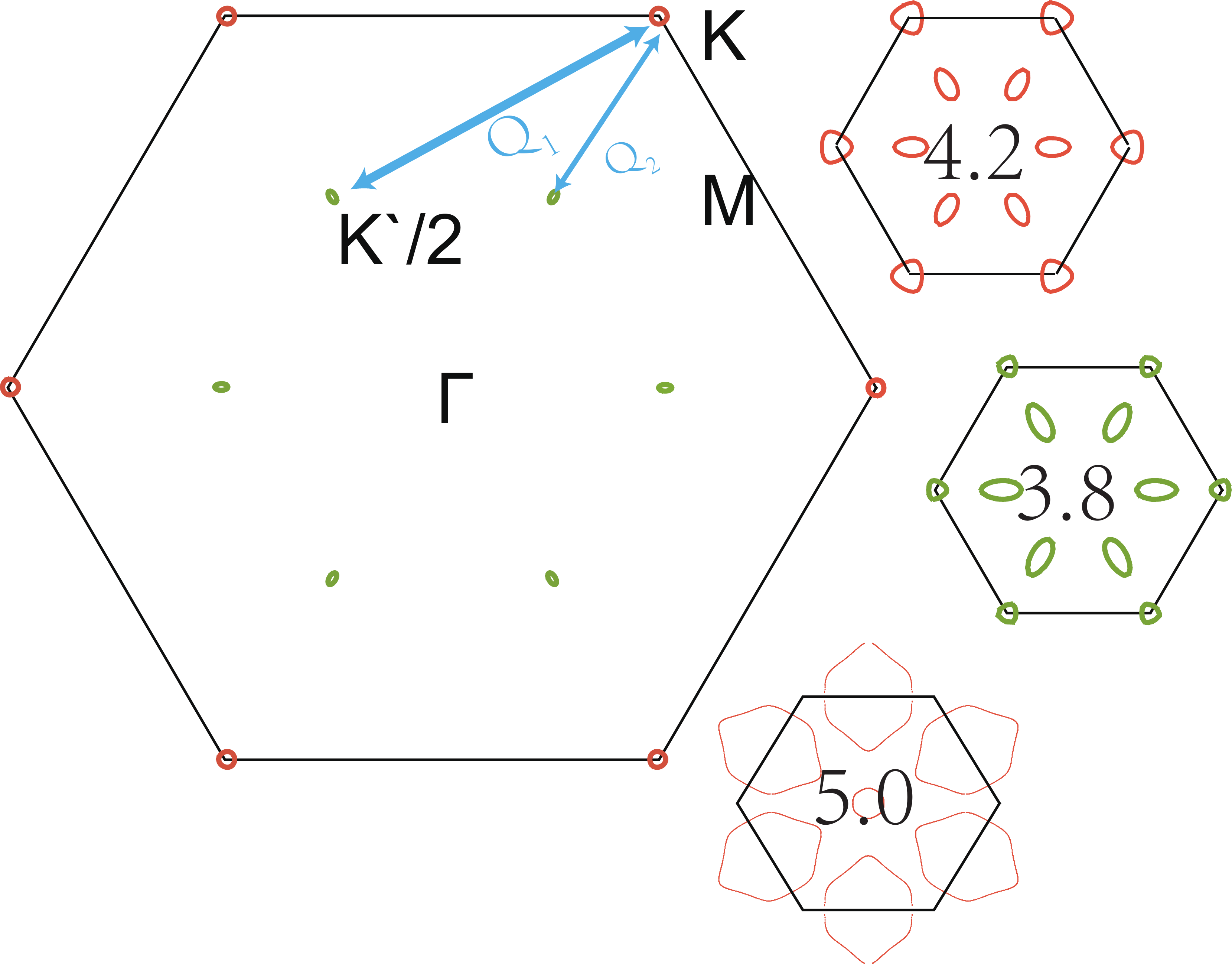}}\caption{
Fermi surfaces and  nesting vectors. The electron pockets and hole pockets are marked by red and green respectively.   The insets from top to bottom are Fermi surfaces at three doping levels , 0.1 (electron), -0.1 (hole) and 0.5 (electron) per Ni atom with respect to the half filling, corresponding to formula $x= (n-4)/2$ with $n$ the total electrons in each unit cell. In (a), nesting vectors $\mathbf{Q}_1$ and $\mathbf{Q}_2$ are depicted.
 \label{Fermi-surface} }
\end{figure}

The dominance of the $t^{TNN}_{yzyz}$ in \ce{NiPX3} can be understood from the lattice chemistry.  The Ni-$e_g$ orbitals are strongly coupled with S/Se-p orbitals.   These effective hoppings  are mediated through the central \ce{P2X6^{4-}} anion. For the NN hoppings, two NN Ni atoms are in two edge shared \ce{MX6} octahedral complexes. As the Ni-X-Ni angle is close to 90\degree, the NN indirect hopping through X is very small. The SNN effective hopping is mediated by two S/Se atoms which separately locate in the top and bottom layers. The coupling between these two S/Se atoms is weak due to the long distance around 3.8\AA  ~between them, which explains the weak SNN   hoppings. By contrast, the TNN $\sigma$ hopping parameter  is mediated through two S/Se atoms in the same layer.

The effects of other hopping parameters on Dirac points and band structures are indicated in Fig.\ref{hopping-tnn} (b), in which the arrows represent the motion of Dirac points and band structures when the corresponding hopping parameters increase.  More specifically,  the weak third neighbor $\pi $-bond   hoppings $t^{TNN}_{xzxz}$ neither affect the Dirac cones at $K$ and $K/2$, nor the band degeneracy points at $\Gamma$ and $M$. They only affect the flat bands  in Fig.\ref{hopping-tnn} (a) far away from Fermi energy.  The flat bands turn to disperse when $t^{TNN}_{xzxz}$ increases.  Therefore, in the weak hopping region, the low energy physics near Fermi surfaces are not affected by the third neighbor $\pi $-bond  hoppings.  The weak second nearest neighbor hoppings, $t^{SNN}_{xzxz}$ and $t^{SNN}_{yzyz}$,  shift the Dirac points at $K/2$ and $K$  vertically. By increasing these hoppings, the two Dirac points shift in opposite directions by a shift ratio equal to $3$ as indicated by the purple arrows in Fig.\ref{hopping-tnn} (b).  The weak NN hoppings, $t^{NN}_{xzxz}$ and $t^{NN}_{yzyz}$, donot affect the Dirac cone at $K$  because of  the symmetry protection from the $C_3$, time reversal and inversion symmetries. However, it drags the $K/2$ Dirac cone along $\Gamma-K$ line as indicated by the green arrows in Fig.\ref{hopping-tnn} (b). Band crossing  points $D_3$ and $D_4$ are also dragged along the direction indicated by   the green arrows in Fig.\ref{nm_band} (b). 

The Fermi surfaces at the different doping levels are shown in Fig.\ref{Fermi-surface}.   Without the SNN hoppings, the model gives Dirac semimetals at half filling. Thus,  the tiny pockets at half filling shown in Fig.\ref{Fermi-surface} stem from very small SNN hoppings.  Due to charge conservation, the area of electron pockets at $K/2$ are three times smaller than those hole pockets at $K$. In principle, with very small hole doping, strong nesting can take place between the electron and hole pockets at $K$ and $K^\prime/2$ respectively, but not  at $K$ and $ K/2$, by taking into consideration of the shapes of Fermi pockets. By increasing hole (electron) doping, both pockets at $K$ and $K/2$ become hole (electron) pockets. When the doping reaches around 0.3 carriers per Ni atoms, there is a Lifshitz transition of Fermi surfaces, namely,  the two pockets emerge together to become one Fermi surface. 

\begin{figure}
\centerline{\includegraphics[width=0.5\textwidth]{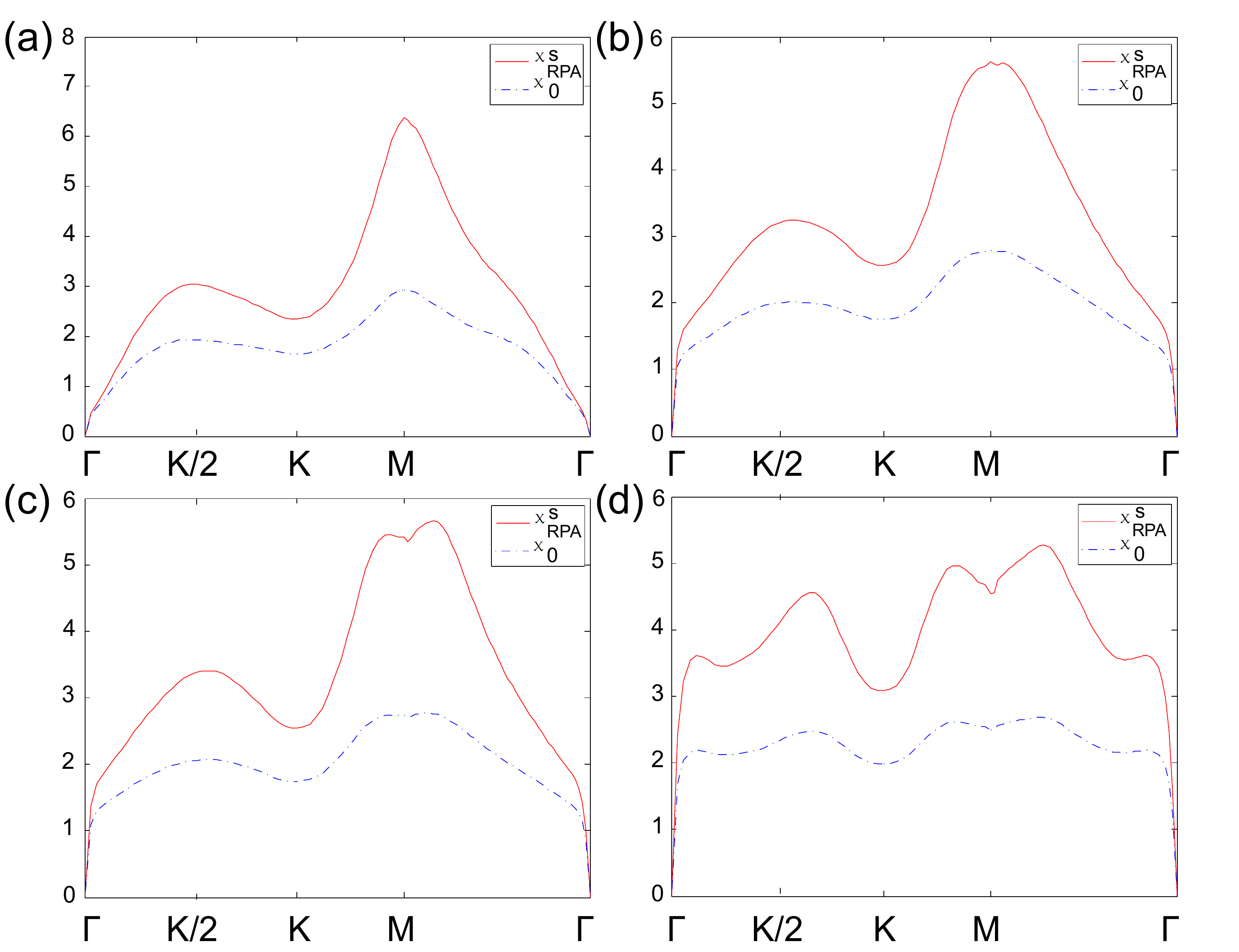}}\caption{
Bare (dashed blue) and RPA (solid red) approximated spin susceptibility   for different doping levels: (a) half filling, (b) 0.1, (c) -0.1 and (d) 0.3 (Liftshiz point where $K$ and $K/2$ pockets connect). Here the onsite energy $U=0.3$eV and Hund's coupling $J_h=0.2U$ is adopted, similar to ref.\cite{Liyinxiang2017}. Resonance apexes appear around   the nesting vector $\mathbf{Q}_1$ (M) and $\mathbf{Q}_2$ (K/2) marked in Fig.\ref{Fermi-surface}.
 \label{spin-sus} }
\end{figure}

From the above Fermi surface topology, we can consider the possible Fermi surface nesting in a paramagnetic state. Near half filling,   the nesting vector is given by $\mathbf{Q}_1=\mathbf{G}/2$,  half of the reciprocal lattice vector, as highlighted in Fig.\ref{Fermi-surface}. This vector is exactly the ordered magnetic wavevector in  the AFM  zigzag state.  We calculate the spin susceptibility under random phase approximation (RPA), with the same method and notations specified in literature\cite{Liyinxiang2017,Scalapino2012}. The result is plotted in Fig.\ref{spin-sus} for several different doping levels. Clearly the susceptibility peak emerges at M ($\mathbf{Q}_1$) near half filling.  Below the critical doping at the Lifshitz transition, the peak is well preserved, indicating the existence of  strong AFM fluctuations.

\section{Magnetic exchange coupling parameters and the AFM Zigzag state}
\label{s3}
Without doping, \ce{MPX3} are known to be magnetic insulators\cite{wang_new_2018,susner_metal_2017,wildes_magnetic_2015,chittari_electronic_2016}. As the magnetic moments   are localized at the transition metal atoms, the magnetism can be captured by an effective Heisenberg model with local magnetic moments. As the effect of the spin orbital coupling is generally small for $e_g$ orbitals, we expect an isotropic Heisenberg model. Furthermore, from the lattice structure, it is obvious that the minimum effective model should include  NN, SNN, and TNN magnetic exchange coupling parameters. Namely,    the model can be written as
 \begin{eqnarray}
 H&=J_1\sum_{<ij>_{NN}} \vec S_i\cdot \vec S_j+J_2\sum_{<ij>_{SNN}} \vec S_i\cdot \vec S_j\nonumber\\
 &+J_3\sum_{<ij>_{TNN}} \vec S_i\cdot \vec S_j.
 \end{eqnarray}

\begin{figure}
\centerline{\includegraphics[width=0.5\textwidth]{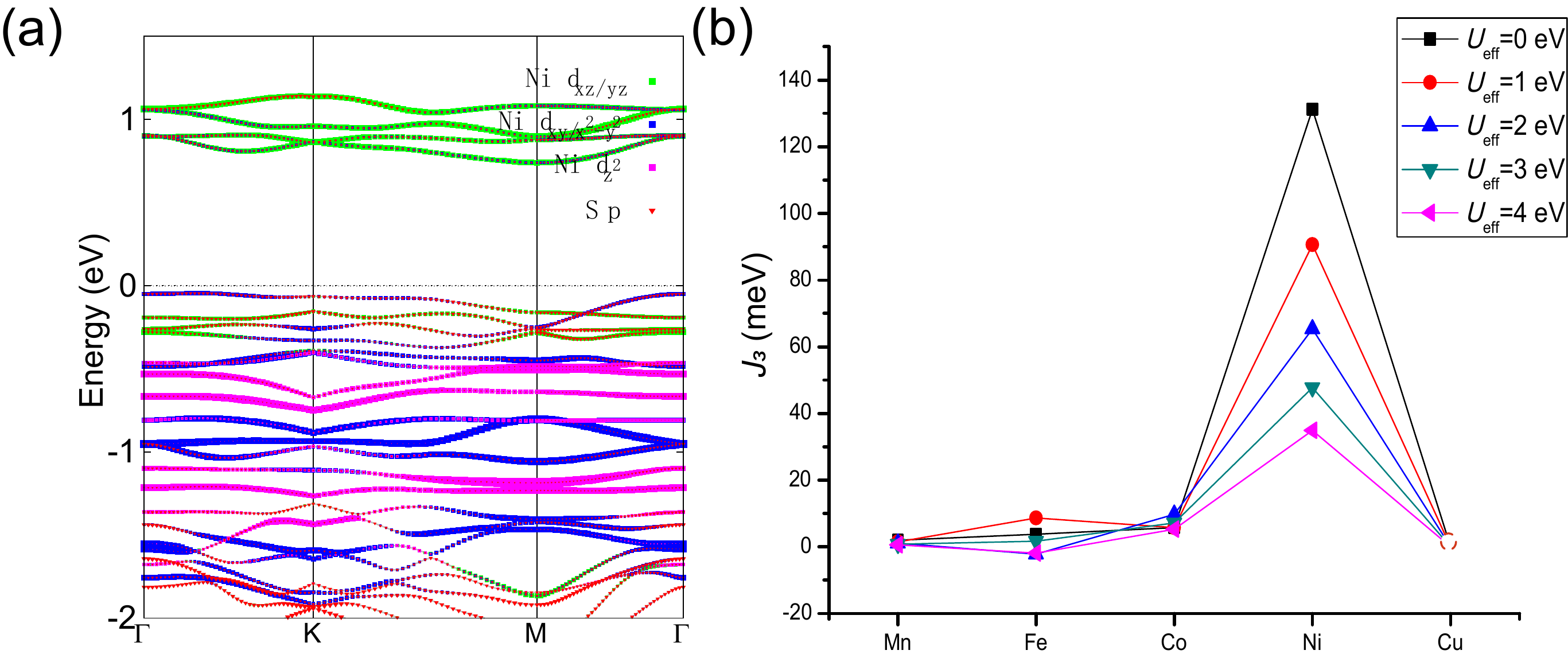}} \caption{(a) The band structure of \ce{NiPS3} in the AFM zigzag state. (b)  $J_3$ superexchange  AFM interactions  in \ce{MPS3}  (M = Mn, Fe, Co, Ni, Cu), which are extracted from the GGA+$U$ calculations with the values $U_{eff}= (0, 1, 2, 3, 4) \text{eV}$.
 \label{mag_J3} }
\end{figure}

 To extract  the magnetic exchange coupling parameters,  we consider the following four different magnetic states: the ferromagnetic (FM) state, the  AFM Neel  state, the  AFM zigzag  and the  AFM stripy for \ce{MPS3} (M=Mn,Fe,Co,Ni,Cu) which   have been synthesized experimentally. Those four magnetic ordering arrangements are shown in the review\cite{wang_new_2018}. The  AFM zigzag state is shown in Fig.\ref{sublattice} (a).    The results are shown in Table.\ref{mag_data}.  We find that the AFM Neel state is favored for \ce{MnPS3} and the AFM zigzag is favored for \ce{FePS3},\ce{CoPS3} and \ce{NiPS3}, which is consistent with  the experimental results in bulk \ce{MPS3} materials\cite{le_flem_magnetic_1982}. Our DFT calculation can give the insulating states  even without considering $U_{eff}$.    With $U_{eff}$ in GGA+$U$ method,  all four monolayer transition-metal phosphorous trisulfides  become AFM insulators, as shown in Table.\ref{mag_data}. As a typical example, we plot the   insulating band structure in the AFM zigzag state for \ce{NiPS3}  in Fig.\ref{mag_J3} (a ). The  Mn, Fe, Co and Ni atoms are in high spin states and the magnetic moments slightly increase as $U_{eff}$ increases.  For bulk materials, the experimental band gaps are 3.0 eV, 1.5 eV and 1.6 eV for Mn,Fe and Ni-based compounds, respectively\cite{wang_new_2018}.  As shown in Table.\ref{mag_data}, the calculated band gaps by GGA+$U$ at $U_{eff}=4\text{eV}$ are quantitatively close to the experimental values. 

\begin{table}
\caption{\label{mag_data}%
The calculated ground state magnetic orders, magnetic moments and the band gaps for monolayer \ce{MPS3} (M = Mn, Fe, Co, Ni) using GGA+$U$ ($U_{eff}=0$ or $4\text{eV}$). }
\begin{ruledtabular}
\begin{tabular}{cccc}
$U_{eff}=0$  &Ground state magnetic order& Moment (\text{$\mu_B$})& Gap (eV) \\
 \colrule
 {\ce{MnPS3}} & AFM Neel   & 4.26 & 1.43\\
 {\ce{FePS3}} & AFM zigzag & 3.32 & 0.01 \\
 {\ce{CoPS3}} & AFM zigzag & 2.24 & 0.02 \\
 {\ce{NiPS3}} & AFM zigzag & 1.13 & 0.79\\
 {\ce{CuPS3}} & FM & 0.24 & 0\\
 \colrule
 $U_{eff}=4eV$  &Ground state magnetic order& Moment (\text{$\mu_B$})& Gap (eV)\\
 \colrule
 {\ce{MnPS3}} & AFM Neel   &4.53&2.39\\
 {\ce{FePS3}} & AFM zigzag &3.61&2.06\\
 {\ce{CoPS3}} & AFM zigzag &2.57&1.85\\
 {\ce{NiPS3}} & AFM zigzag &1.44&1.66\\
 {\ce{CuPS3}} & FM         & 0.24 & 0
\end{tabular}
\end{ruledtabular}

\end{table}

The classical energies of the four different magnetic states for the effective Heisenberg model are given by
\begin{eqnarray}
& E_{FM}&=S^2 (6J_1+12J_2+6J_3)+E_0,\nonumber \\
&E_{AFM-Neel}&=S^2 (-6J_1+12J_2-6J_3)+E_0,\nonumber \\
&E_{AFM-zigzag}&=S^2 (2J_1-4J_2-6J_3)+E_0,\nonumber \\
&E_{AFM-stripy}&=S^2 (-2J_1-4J_2+6J_3)+E_0
\end{eqnarray}
From the calculated energies of these states, we can extract the effective magnetic exchange interactions. The results are listed in Table.\ref{J_value}. Some similar results have been obtained previously\cite{wildes_magnetic_2015,chittari_electronic_2016}. Our calculation  are consistent with these previous calculations\cite{wildes_magnetic_2015,chittari_electronic_2016}. 

Here we pay special attention to the values in \ce{NiPX3}. As shown in Table.\ref{J_value}, for \ce{NiPX3}, among the three magnetic exchange coupling parameters, $J_3$  is one order of magnitude larger than the  other two parameters. Moreover, $J_3$ is strongly AFM while $J_1$ and $J_2$ both are weakly FM. These qualitative  features are independent of $U_{eff}$. The dominance of $J_3$ over the other two further confirms the extracted physical picture of weakly coupled four sublattices as shown in Fig.\ref{sublattice} (b)  based on  the hopping parameters in the electronic band structure.

$J_3$ stems from  so-called AFM super-superexchange interaction\cite{le_flem_magnetic_1982,wildes_magnetic_2015}. In   Fig.\ref{mag_J3} (b), we plot the values of $J_3$ as a function of M (M=Mn, Fe, Co, Ni, Cu). It is important to note that for \ce{CuPS3}, the AFM Neel and AFM zigzag states are not metastable with different $U_{eff}$ in our calculation, which means that $J_3\le0$. In Fig.\ref{mag_J3} (b),  it is clear that $J_3$ reaches the maximum value in  \ce{NiPX3}, which can be easily understood as  the hall-filling of $e_g$ orbitals maximize the super-superexchange interaction.

\begin{table}
\caption{\label{J_value}
The calculated exchange interaction parameters $J_1$, $J_2$ and $J_3$ for monolayer \ce{MPS3} (M = Mn, Fe, Co, Ni) and \ce{NiPSe3} using GGA+$U$ ($U_{eff}=0$ or $4\text{eV}$).}
\begin{ruledtabular}
\begin{tabular}{cccc}
 $U_{eff}=0$  & $J_1$ (meV) & $J_2$ (meV) & $J_3$ (meV) \\
 \colrule
 {\ce{MnPS3}} & 4.69  &	0.40 &	2.03 \\
 {\ce{FePS3}} & -20.90 & 5.08 &	3.81 \\
 {\ce{CoPS3}} & -22.52 & 14.50 & 5.78 \\
 {\ce{NiPS3}} & -10.63 & -2.30 & 131.24 \\
 {\ce{NiPSe3}}& -20.52 &-4.38 & 215.55 \\
 \colrule
 $U_{eff}=4eV$  & $J_1$ (meV) & $J_2$ (meV) & $J_3$ (meV)\\
 \colrule
 {\ce{MnPS3}} & 1.62 &	0.09 &	0.61 \\
 {\ce{FePS3}} & -6.39&	4.11 &	-1.90 \\
 {\ce{CoPS3}} &-1.71 & -0.07 & 5.23  \\
 {\ce{NiPS3}} &-5.17 & -0.78 & 34.93 \\
 {\ce{NiPSe3}}&-5.80 & -0.45 & 53.13 \\
\end{tabular}
\end{ruledtabular}

\end{table}

\section{The two-orbital t-J model and doping phase diagram for  \ce{NiPX3}}
\label{s4}
\begin{figure}[ht]
\centering
\includegraphics[width=0.5\textwidth]{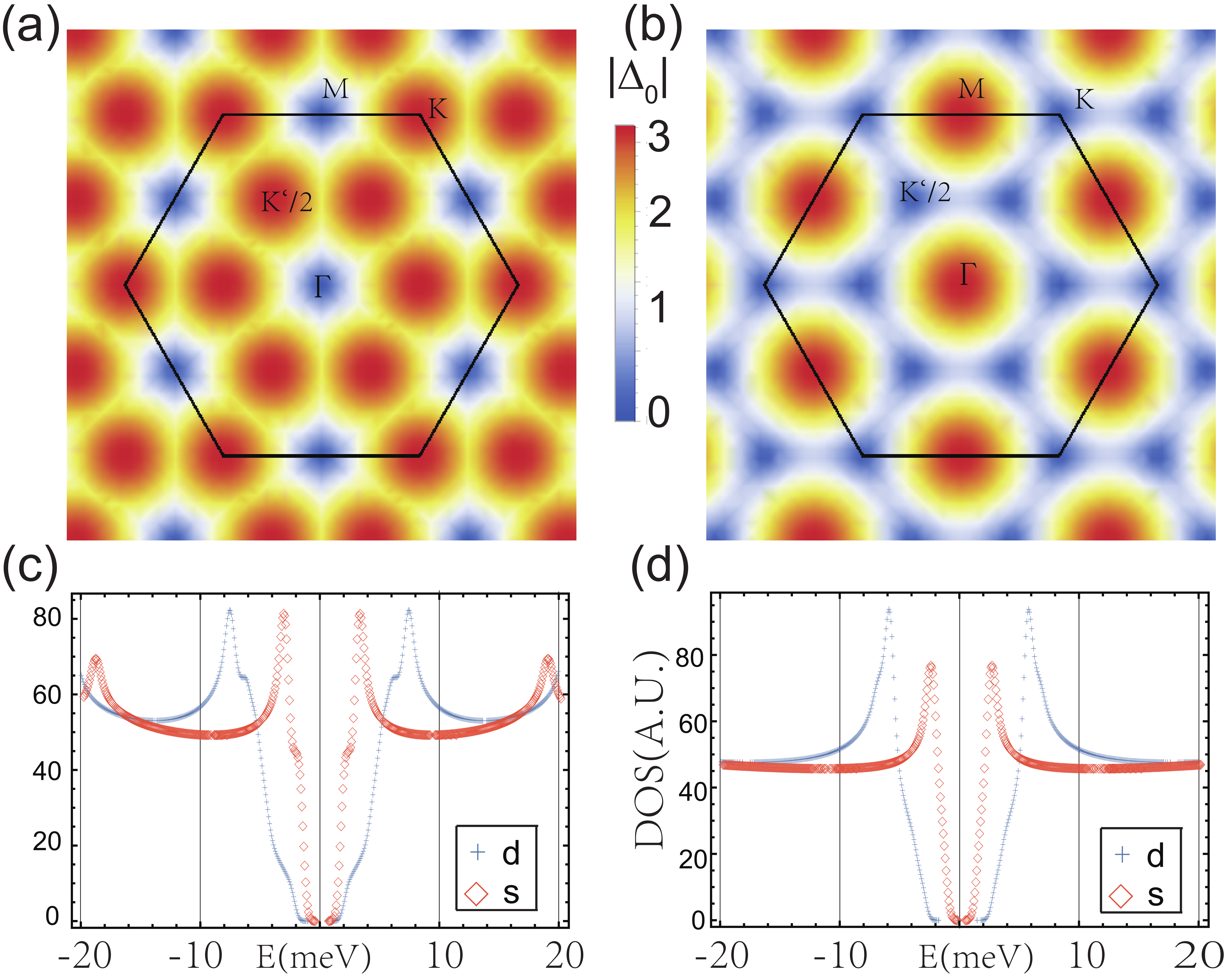}
\caption{(a) and (b)  plot the amplitude distributions of the superconducting gap $G(k)$ of the $d\pm id$-wave  and the extended $s$-wave states respectively.  (c) and (d) are the density of states in the $d\pm id$-wave  and the extended $s$-wave states at a doping  level, n=3.8 and 4.2, respectively, by taking $\Delta_0=0.3$ and $J_3=40meV$.  }\label{fig:pairing}
\end{figure}

From the above analysis and the known experimental facts\cite{le_flem_magnetic_1982}, it is clear that 
\ce{NiPX3} must belong to strongly correlated electron systems.  The band width of the two $e_g$ orbitals is only about 1eV, much less than the band gaps in their AFM zigzag states.   Moreover, as we showed above, the experimental band gaps are close to the theoretical results when we take $U_{eff}\sim 4eV$, which is much larger than the band width as well.  Thus, the magnetic order is caused by the strong electron-electron correlation.

Following the standard argument, \ce{NiPX3}, just like many other strongly correlated electron systems,   must be a Mott insulator. As  $U_{eff}$ is much larger than the band width, we can take the large U limit to derive a t-J type of model.  For \ce{NiPX3},   a minimum two-orbital t-J model can be written as
\begin{equation}
    H_{tJ}=\hat P H_0\hat P +\sum_{ij} H_{J,<ij>},
    \label{htj}
\end{equation}
where $\hat P$ is the projection operator to remove the double occupancy and $H_{J,<ij>}$ is the effective interaction.  In a two orbital model, in general, we should consider  a spin-orbital Kugel-Khomskii type of superexchange interactions\cite{castellani-prb78,kk}. However, here because of the hopping is dominated by the TNN $\sigma$ couplings, in the first order approximation,  the leading interaction can be derived as
\begin{eqnarray}
H_{\rm J,<ij>}=  J_{3} ({\bf S}_{i,\bar{ij}}\cdot{\bf S}_{j,\bar{ij}}-\frac{1}{4}{\bf n}_{i,\bar{ij}}{\bf n}_{j,\bar{ij}}),
\end{eqnarray}
where $<ij>$ is a TNN link,  ${\bf S}_{i,\bar{ij}}$ and ${\bf n}_{i,\bar{ij}}$ are the spin and density operators of the electron located at the orbital which participates $\sigma$ hopping through the $<ij>$ link at the ith site respectively. 

Before we present a full mean field calculation for the above model, we would like to qualitatively argue possible superconducting states. By decoupling the $J_3$ AFM interaction in  the pairing channel, the Bogoliubov-de Gennes (BdG) Hamiltonian in Nambu space $\Psi^\dagger_k= (\psi^\dagger_{k\uparrow},\psi^T_{\bar{k}\downarrow})$ for a uniform superconducting state can be generally written as 
\begin{align}
h_{Bk}=\left (
\begin{array}{cc}
h_k & \Gamma_k\\
\Gamma^\dagger_k & -h^{T}_{\bar{k}}
\end{array}\right),\qquad 
\Gamma_k=\left (
\begin{array}{cc}
0 & \Delta_k\\
\Delta^{T}_{\bar{k}} & 0
\end{array}\right),\label{equ:bdg}
\end{align}
with $\bar{k}\equiv -k$ and $h_k$ given in Equ.\ref{hk}. Here  the general form of pairing matrix is 
\begin{align}
\Delta_k=\sum_{j} e^{-i\mathbf{k}\cdot \mathbf{a_3}_j}\Delta_j e^{i l \theta_j},\label{equ:pairing}
\end{align}
where $\theta_j$ is the angle of the TNN vector $\mathbf{a}_{3j}$, $\Delta_j$ is the two-orbital pairing matrix on bonds  connected by $\mathbf{a}_{3j}$ and $l$ is the angular momentum quantum number of the order parameter, with ($l=0,\pm 2\cdots$) representing ($s,d\cdots$) waves in the singlet pairing channels. 

If we only consider the  $\sigma$ bond hoppings, the electronic structure is identical to an isotropic one-orbital honeycomb model.   In this case the real space pairing matrices reduce to a constant number $\Delta_j \sim \Delta_0$ and the true gap can be represented with $ G (k)=\sqrt{\max (|\Delta_k|^2,|\Delta_{\bar{k}}|^2)}$.  The  $\Delta_k$ for the $d\pm id$-wave and the extended $s$ wave  can be explicitly written as 
\begin{eqnarray}
\Delta^s_k&=& \Delta_0 (e^{2ik_y}+2e^{-ik_y}\cos{ (\sqrt{3}k_x)}) \\
\Delta^{d\pm id}_k&=&-\Delta_0 (e^{2ik_y}+2 e^{-ik_y}\cos{ (\sqrt{3}k_x\pm \frac{2\pi}{3})})
\end{eqnarray}
In Fig.\ref{fig:pairing} (a,b), we  draw the one-orbital $d\pm id$-wave and the extended $s$ wave gap distribution. The  $d\pm id$ pairing gap peaks locate at $K(K^\prime)$ and $K(K^\prime)/2$ while the $s$ pairing gap has peaks around $\Gamma$ and $M$. At low doping, the Fermi surfaces  are around $K(K^\prime)$ and $K(K^\prime)/2$ as seen from Fig.\ref{nm_band}. Thus, following the general argument given in\cite{hu_local_2012}, known as the Hu-Ding principle,  the $d\pm id$-wave pairing is favored over the $s$-wave pairing as the former would open much bigger superconducting gap on the Fermi surfaces to save more energy than the latter. 

Although the analytic formula for the gap can not be obtained, this  above analysis can be extended to the two orbital model. In Fig.\ref{fig:pairing} (c,d), we numerically calculate the density of states of the $d\pm id$-wave and $s$-wave superconducting states in the full two-orbital model by assuming the pairing amplitudes $\Delta_0$ in both states are identical in all the $\sigma$ bonds.  It is clear that the $d\pm id$ state has much bigger superconducting gap than the $s$-wave state at low doping. Moreover, The $s$-wave pairing state can be gapless while the $d\pm id$ state has a full gap, which stems from the mismatch of the $s$-wave pairing momentum form factor with the normal state Fermi surfaces, the essential idea behind the Hu-Ding principle\cite{hu_local_2012}. With much large doping level,  the Lifishitz transition of the Fermi surfaces  merges pockets around $M$ and $\Gamma$.  As a result, the $s$ wave pairing can become highly competitive. However, in this region, the AFM fluctuation also becomes very weak so that the superconductivity likely vanishes. 

\begin{figure}[ht]
\centering
\includegraphics[width=0.5\textwidth]{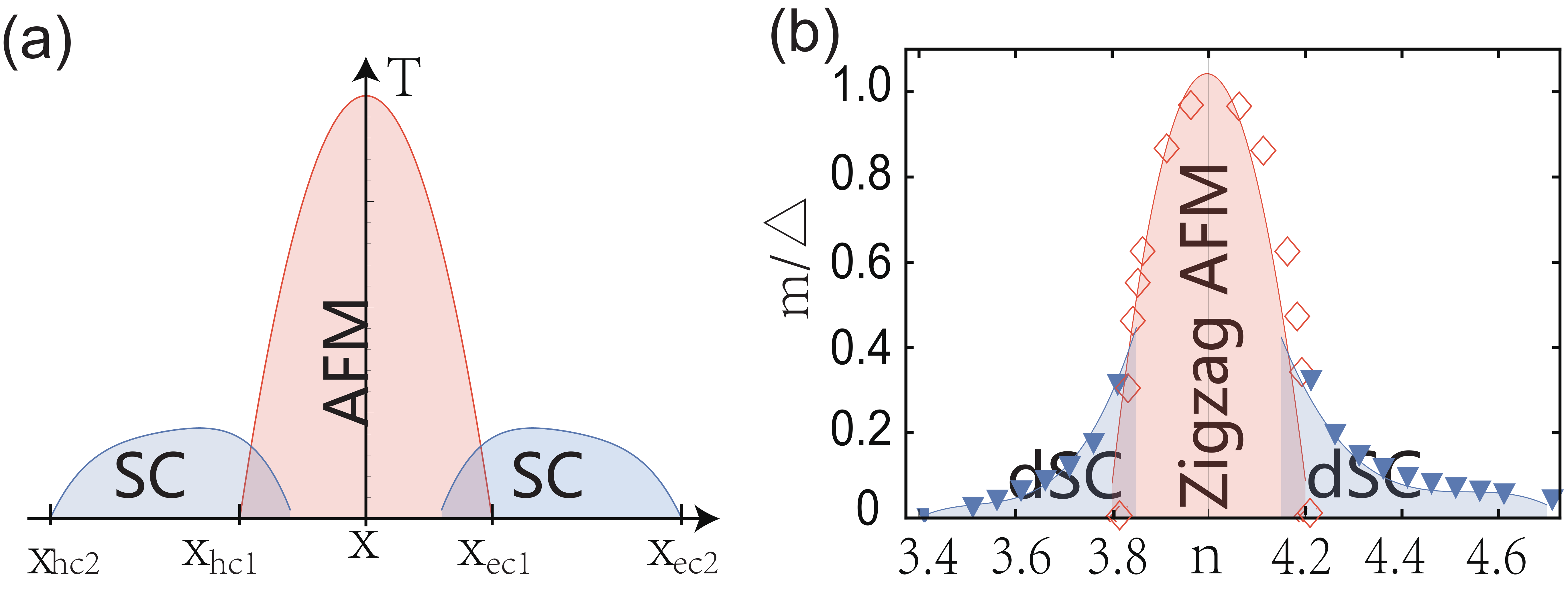}
\caption{The phase diagram under doping.  (a) The  sketch of a typical magnetism versus superconductivity phase diagram.  (b) The Zigzag magnetic  order parameter and $d\pm id$ pairing strength as doping increases in the two-orbital model calculated by the slave-Boson meanfield with $J_3=40meV$. When the electrons per unit cell $n=4.2$ ($x_{c1}=0.1$), zigzag antiferromagnetic order parameter vanishes. As the doping increases,  the superconductivity order parameter  decreases and vanish around $n=4.7$ ($x_{c2}=0.35$). The hole doped phase diagram is almost symmetric for magnetic phase, while superconductivity vanished around $n=3.4$ (0.3 hole doping).  }\label{fig:phase}
\end{figure}

Our slave-boson meanfield result on the pairing symmetry is consistent with above analysis.  Here we  report the magnetic and superconducting phase diagram from the U (1) slave-boson meanfield for the  model in Eq.\ref{htj} \cite{QHWang-prb04,PatrickLee01renormalized}. The method has been shown to provide correct qualitative information of the phase diagram as a function of doping.  The slave-boson measures the electron occupancy and leads to the renormalization of hopping amplitude\cite{Sigrist05slave}. The detailed procedure is given in Appendix B. An  illustration of the phase diagram is sketched in Fig.\ref{fig:phase} (a) with the doping vs temperature. At low doping $x<x_{c1}$, the system would stay in antiferromagnetic state. Between two critical doping levels $x_{c1}<x<x_{c2}$, it is the superconducting phase. Near the quantum critical point, there might be coexistence of magnetism  and superconductivity, or some other rich intertwined orders. 

Our meanfield calculation results are plotted in Fig.\ref{fig:phase} (b). Here, we adopt $J_3=40meV\sim t^{TNN}_{yzyz}/6$. The density $n=4$ represents half-filling and the doping level of $x$ electron per Ni atom is corresponding to $n= 4+2x$. Owing to the orbital selective exchange, the long range  AFM zigzag   order vanishes around the doping level of 0.1 electron per Ni atom (n=4.2) away from half-filling.  If we take $J_3=80$ meV, this critical doping value increases 0.2 per Ni atom (n=4.4), which is similar to the meanfield result in cuprates\cite{QHWang-prb04} by the same method.  Superconducting order parameters also decreases as doping increase. When the doping reaches 0.35 per Ni atom (n=4.7),  the superconducting order parameter $\Delta_0$ becomes too small to have any physical meaning.  In all these doping region with superconductivity,   $d\pm id$ wave pairing is energetically favorable over the extended $s$-wave. It is worthy  mentioning that the phase diagram is slightly asymmetric between the hole and electron doped region. The magnetism is almost symmetric, vanishing around $x_{c1}\approx \pm 0.1$ and the superconductivity decreases slightly faster with hole doping.

\section{Discussion}
In summary, we have shown that the Ni-based TMTs are close to a strongly correlated quadruple-layer graphene and are Dirac materials described by a two-orbital model with the strong electron electron interaction.  The main electronic kinematics and magnetic interactions exist with unusual long range distance between two third nearest neighbour Ni atoms, which stems from the super-super exchange mechanism. With this underlining  electronic structure,  the materials  provide a simple and ideal  playground to investigate strong correlation physics.

 The two orbital model can be viewed as a natural extension of the single orbital model in the conventional high temperature superconductors, cuprates.  Recently, materials with  both active $e_{g}$ orbitals  have gained  much attention. Ba$_2$CuO$_{3+\delta}$\cite{li2018new}, synthesized under high pressure, is likely an extremely heavy hole doped cuprates. As the Jahn-Teller  distortion  of the CuO$_6$ octahedron causes a shorter Cu-O bond along c-axis  than  the in-plane ones, both $e_{g}$ orbitals become important\cite{maier2018d}.  The single CuO$_2$ layer grown by MBE  also has a similar electronic structure\cite{jiang2018nodeless}. In La$_2$Ni$_2$Se$_2$O$_3$, a recent theoretically proposed candidate of Ni-based high temperature superconductors\cite{le_possible_2017},  the low energy physics is also attributed to the two $e_g$ orbitals of Ni atoms.  If  the superconductivity is determined to arise  in the two orbital model, the Ni-based TMTs, together with these materials,  can provide us much needed information to solve the elusive mechanism of high temperature superconductivity. 

Comparing with recent artificial graphene systems, in which  new bands are created to reduce the kinetic energy  so that  the effect of the weak electron-electron correlation can arise in the standard graphene\cite{cao2018correlated}, the Ni-based TMTs are simply in the other limit (Mott limit)  with the strong electron-electron correlation. We can  consider to increase the kinetic energy, namely the band width, to enter the Mott transition critical region.  In general, the band width can be increased by applying pressure or by atom substitutions. During these processes,  the angles between Ni-S/Se-Ni can be changed greatly as well, which can lead to different intriguing physics. For example, in this case, more spin-orbital superexchange interaction terms may be important, which can lead to exciting interplay between orbital and spin degrees of freedom. 

Experimentally, electron or hole carriers have not been introduced to Ni-based TMTs. Recently, modern gating technology can induce carriers to a variety of two dimensional materials\cite{Ye2009}. The Ni-based TMTs can be an important playground for this modern technology.

{\it Acknowledgement:} We thank Hong Jiang, Xinzheng Li, Xianxin Wu, Yuechao Wang, Shengshan Qin and Dong Luan for useful discussion. The work is supported by the Ministry of Science and Technology of China 973 program (Grant No. 2015CB921300, No.~2017YFA0303100, No. 2017YFA0302900), National Science Foundation of China (Grant No. NSFC-11334012), and   the Strategic Priority Research Program of  CAS (Grant No. XDB07000000). Q.Z. acknowledges the support from the International Young Scientist Fellowship of Institute of Physics CAS (Grant No. 2017002) and the Postdoctoral International Program (2017) from China Postdoctoral Science Foundation.

\appendix
\section{The explicit form of effective Hamiltonian }
The tight-binding effective Hamiltonian  $H_0$ in Eq.\ref{h0} is a $4\times 4 $ matrix. The explicit form of its elements is given by
\begin{widetext}
\begin{eqnarray}
H_{11}&=&t^{SNN}_{xzxz} (2\cos{k_x}+\cos{\frac{k_x}{2}}\cos{\frac{\sqrt{3}k_y}{2}})+t^{SNN}_{yzyz}\cos{\frac{k_x}{2}}\cos{\frac{\sqrt{3}k_y}{2}},\nonumber \\
H_{12}&=&t^{SNN}_{xzxz} (-\sqrt{3}\sin{\frac{k_x}{2}}\sin{\frac{\sqrt{3}k_y}{2}})+t^{SNN}_{yzyz} (\sqrt{3}\sin{\frac{k_x}{2}}\sin{\frac{\sqrt{3}k_y}{2}})+i (4t^{SNN}_{xzyz}\sin{\frac{k_x}{2}} (\cos{\frac{k_x}{2}}-\cos{\frac{\sqrt{3}k_y}{2}})),\nonumber \\
H_{13}&=&\frac{1}{2} e^{-\frac{2 i {k_y}}{\sqrt{3}}} \left (e^{\frac{1}{2} i \sqrt{3} {k_y}} \cos \left (\frac{k_x}{2}\right) ({t^{NN}_{xzxz}}+3 {t^{NN}_{yzyz}})+e^{i \sqrt{3} {k_y}} (\cos ({k_x}) ({t^{TNN}_{xzxz}}+3 {t^{TNN}_{yzyz}})+2 {t^{NN}_{xzxz}})+2 {t^{TNN}_{xzxz}}\right), \nonumber \\
H_{14}&=&\frac{1}{2} i \sqrt{3} e^{-\frac{i {k_y}}{2 \sqrt{3}}} \left (\sin \left (\frac{{k_x}}{2}\right) ({t^{NN}_{xzxz}}-{t^{NN}_{yzyz}})-e^{\frac{1}{2} i \sqrt{3} {k_y}} \sin ({k_x}) ({t^{TNN}_{xzxz}}-{t^{TNN}_{yzyz}})\right), \nonumber \\
H_{22}&=&t^{SNN}_{yzyz} (2\cos{k_x}+\cos{\frac{k_x}{2}}\cos{\frac{\sqrt{3}k_y}{2}})+t^{SNN}_{xzxz}\cos{\frac{k_x}{2}}\cos{\frac{\sqrt{3}k_y}{2}},\nonumber \\
H_{24}&=&\frac{1}{2} e^{-\frac{2 i {k_y}}{\sqrt{3}}} \left (e^{\frac{1}{2} i \sqrt{3} {k_y}} \cos \left (\frac{{k_x}}{2}\right) (3 {t^{NN}_{xzxz}}+{t^{NN}_{yzyz}})+e^{i \sqrt{3} {k_y}} (\cos ({k_x}) (3 {t^{TNN}_{xzxz}}+{t^{TNN}_{yzyz}})+2 {t^{NN}_{yzyz}})+2 {t^{TNN}_{yzyz}}\right),\nonumber \\
\end{eqnarray}
\end{widetext}
with $ H_{23}=H_{14}, 
H_{33}=H_{11}, 
H_{34}=H^*_{12}$ and 
H$_{44}=H_{22}$ by symmetry.  These hopping parameters are given in Table.\ref{hopping-tnn} in the main text.

\section{Formulation of the slave boson meanfield}

We provide the detailed procedure for the slave boson meanfield method on the Hamiltonian Eq. (\ref{htj}). In our  two orbital model, the two $e_g$ orbitals are degenerate so that they have the identical occupancy.   In the slave-boson approximation\cite{Sigrist05slave}, the same occupancy for all the  orbitals leads to the same renormalization for all the hopping interaction. Namely, we have
\begin{eqnarray}
\hat{P}H_0\hat{P}=|\frac{n}{4}-1|H_0,
\end{eqnarray}
in which  $n=4\pm 2x$, $x$ is the doped electron (+) or hole (-) per atom.  $n=4$  represents the half filling, where the kinetic energy vanishes and the Hamiltonian reduces to pure Heisenberg exchange interaction. The exchange term $H_{J,\langle ij\rangle}$ in Eq. (\ref{htj}) can be decoupled in
superconducting and magnetic channels.

 In superconducting channel, it is 
\begin{align}
-\frac{J_3}2\sum_{ij}\left (\langle \Delta^{\dagger}_{ij}\rangle \Delta_{ij}+ \Delta^{\dagger}_{ij}\langle \Delta_{ij}\rangle \right)+E_s,
\end{align}
 with pairing matrix $ \Delta^{\dagger}_{ij}\equiv   \sigma a^\dagger_{i\bar{ij},\sigma}a^\dagger_{j\bar{ij},\bar{\sigma}} $ with $a_{i\bar{ij},\sigma}=\cos\theta_{ij} a_{ix,\sigma}+\sin\theta_{ij} a_{iy,\sigma}$ and the constant part
 \begin{align}
     E_s=&\frac{J_3}{2}\sum_{ij}\left (\langle \Delta^{\dagger}_{ij}\rangle \langle \Delta_{ij}\rangle +n_{i\bar{ij},\sigma}n_{j\bar{ij},\sigma}\right),
 \end{align}
with the spin indices $\bar{\sigma}=-\sigma=\pm 1$ and link $ij$ the TNN bond.   The same spin coupling density term $n_{i\bar{ij},\sigma}n_{j\bar{ij},\sigma}$ term in $E_s$ would be decoupled as  bond hopping   $\langle a^\dagger_{i\bar{ij},\sigma} a_{j\bar{ij},\sigma}\rangle a_{i\bar{ij},\sigma} a^\dagger_{j\bar{ij},\sigma}$ according to the approach in literature\cite{QHWang-prb04}. Those bond hopping terms  effectively renormalize the third neighbor $\sigma$-bond hoppings to affect the position and shape of Fermi surface.  The hopping parameters and Fermi surfaces  will be renormalized as the doping varies in our model. Thus as a simple illustrating  of the phase diagram under slave boson method, those bond hopping terms would be simply taken as the density correlation.  Put the renormalized hopping and decoupled exchange terms together, in the two orbital momentum space, we obtain $\sum_{k}\Psi^\dagger_k h_{Bk}\Psi_k+E_s$ with $h_{Bk}$ given in Eq. (\ref{equ:bdg}). 

In the magnetic channel, the $J_3$ dominated spin exchange interaction prefers the  zigzag AFM  state as shown in Fig.\ref{sublattice}. In the mean field level, the magnetic order in the zigzag pattern is given by $<\mathbf{S}_i> = (-1)^{i_2} m_0 \hat{z} $ with $\mathbf{i}=i_1\mathbf{a_2}_1+i_2\mathbf{a_2}_2$. It is important to point out that as only the $\sigma$ bond orbitals are considered, the spin exchange is     orbital selective.  That is to say, the spin operator,  $\mathbf{S}_{i,\bar{ij}}=\frac{1}{2}a^+_{i\bar{ij}\mu}\bm{\sigma}_{\mu\nu}a_{i\bar{ij}\nu}$ with $\bm{\sigma}$ the vector of three Pauli matrices. As a result
\begin{align}
    \langle S^z_{i,\bar{ij}}\rangle =&\cos^2\theta_{\bar{ij}} m_{ixx} +\sin^2\theta_{\bar{ij}} m_{iyy}\\\nonumber
&+\cos{\theta_{\bar{ij}}}\sin{\theta_{\bar{ij}}} (m_{ixy}+m_{iyx})
   ,
\end{align}
with  $m_{i\alpha\beta}\equiv \langle \sigma (a^\dagger_{i\alpha\sigma}a_{i\beta\sigma}))\rangle /2= (-1)^{i_2} m_{\alpha\beta}$.  Up to a constant term, $H_{J,\langle ij\rangle}$ is decoupled as 
\begin{align}
\sum H_{J,\langle ij\rangle}= \sum _k \sigma a^\dagger_{k\alpha \sigma }M_{2\alpha\beta} a_{k+Q_1\beta \sigma}
\end{align}
with   $\mathbf{Q}_1=\mathbf{G}_2/2$  marked in Fig.\ref{Fermi-surface}  as the ordered magnetic wavevector. The scattering matrix  from  $\mathbf{k}$ to $\mathbf{k+Q_1}$ in $e_g$ orbitals space is 
\begin{align}
    M_2=-\frac{3J_3}{16}\left (
\begin{array}{cc}
3m_{xx}+m_{yy} & m_{xy}+m_{yx}\\
m_{xy}+m_{yx} & m_{xx}+3m_{yy}
\end{array}\right).
\end{align}
Defining $\Psi^\dagger_{Mk}= (\Psi^\dagger_k,\Psi^\dagger_{k+Q_1})$,  the meanfield  Hamiltonian can be written as 
\begin{align}
H=\sum_{k\in rBZ} \left (\Psi^\dagger_{M}A_{k}\Psi_{Mk}+tr (h_k+h_{k+Q_1})-8\mu \right)+E_{sm},
\end{align}
with rBZ represent the reduced Brillouin zone due to magnetic cell and $A_k$ is a $16\times 16$ matrix as
\begin{align}
A_{k}=&\left (
\begin{array}{cc}
h_{Bk} & I_2 \bigotimes I_2\bigotimes M_2\\
I_2\bigotimes I_2\bigotimes M_2^\dagger & h_{Bk+Q}
\end{array}\right)\\
\frac{E_{sm}}{N}=&\frac{3J_3}{2}\Delta^2_0 +\frac{3J_3}{8} (3m_{xx}^2+3m_{yy}^2+2m_{xx}m_{yy}\\\nonumber
    &+ (m_{xy}+m_{yx})^2+n^2).
\end{align}
In $A_k$, the first $I_2$ is for particle-hole space and the second is for A-B sublattice. The $h_{Bk}$ is given in Eq. (\ref{equ:bdg}).
 The self-consistency of the chemical potential is also taken into consideration for a fixed doping. It is easy to show that 
\begin{align}
n_k+n_{k+Q_1}=&8+tr (\Sigma \langle \Psi_{Mk} \Psi^\dagger_{Mk} \rangle )\\\nonumber
        =&8-tr (\Sigma U_kf (\Lambda_k)U_k^\dagger),
\end{align}
in which $\Sigma$ is  the $16 \times 16 $  stagger matrix $\Sigma\equiv -I_2\bigotimes \sigma_3\bigotimes I_4$,  $U^\dagger_k A_k U_k=\Lambda_k$, diagonalizes the $A_k$ and $f (\Lambda_k)$ is  the Fermi distribution function.

  It is worthy to mention that the numerical result indicates slight difference between the intra-orbital magnetic orders, $m_{xx}$ and $m_{yy}$, due to the rotation symmetry breaking in the AFM zigzag state. The inter-orbital magnetic orders, $m_{xy}$ and $m_{yx}$, are very small and can be ignored in the meanfield solution.


\begin{thebibliography}{10}
\expandafter\ifx\csname url\endcsname\relax
  \def\url#1{\texttt{#1}}\fi
\expandafter\ifx\csname urlprefix\endcsname\relax\def\urlprefix{URL }\fi
\providecommand{\bibinfo}[2]{#2}
\providecommand{\eprint}[2][]{\url{#2}}

\bibitem{graphene2004}
\bibinfo{author}{Novoselov, K.~S.} \emph{et~al.}
\newblock \bibinfo{title}{Electric field effect in atomically thin carbon
  films}.
\newblock \emph{\bibinfo{journal}{Science}} \textbf{\bibinfo{volume}{306}},
  \bibinfo{pages}{666--669} (\bibinfo{year}{2004}).

\bibitem{wang_new_2018}
\bibinfo{author}{Wang, F.} \emph{et~al.}
\newblock \bibinfo{title}{New frontiers on van der waals layered metal
  phosphorous trichalcogenides}.
\newblock \emph{\bibinfo{journal}{Adv. Funct. Mat.}}
  \textbf{\bibinfo{volume}{28}}, \bibinfo{pages}{1802151}
  (\bibinfo{year}{2018}).

\bibitem{susner_metal_2017}
\bibinfo{author}{Susner, M.~A.}, \bibinfo{author}{Chyasnavichyus, M.},
  \bibinfo{author}{McGuire, M.~A.}, \bibinfo{author}{Ganesh, P.} \&
  \bibinfo{author}{Maksymovych, P.}
\newblock \bibinfo{title}{Metal thio-and selenophosphates as multifunctional
  van der waals layered materials}.
\newblock \emph{\bibinfo{journal}{Adv. Mater.}}
  \textbf{\bibinfo{volume}{29}}, \bibinfo{pages}{1602852}
  (\bibinfo{year}{2017}).

\bibitem{friedel1894thiohypophosphates}
\bibinfo{author}{Friedel, C.}
\newblock \emph{\bibinfo{journal}{Compt. Rend.}} \textbf{\bibinfo{volume}{119}}
  (\bibinfo{year}{1894}).

\bibitem{ferrand1895bull}
\bibinfo{author}{Ferrand, L.}
\newblock \emph{\bibinfo{journal}{Bull. Soc. Chim.}}
  \textbf{\bibinfo{volume}{13}}, \bibinfo{pages}{115} (\bibinfo{year}{1895}).

\bibitem{le_flem_magnetic_1982}
\bibinfo{author}{Le~Flem, G.}, \bibinfo{author}{Brec, R.},
  \bibinfo{author}{Ouvard, G.}, \bibinfo{author}{Louisy, A.} \&
  \bibinfo{author}{Segransan, P.}
\newblock \bibinfo{title}{Magnetic interactions in the layer compounds {MPX}$_3$ ({M = Mn, Fe, Ni; X= X, Se})}.
\newblock \emph{\bibinfo{journal}{J. Phys. Chem. Sol.}}
  \textbf{\bibinfo{volume}{43}}, \bibinfo{pages}{455--461}
  (\bibinfo{year}{1982}).

\bibitem{wang_emergent_2018}
\bibinfo{author}{Wang, Y.} \emph{et~al.}
\newblock \bibinfo{title}{Emergent superconductivity in an iron-based honeycomb
  lattice initiated by pressure-driven spin-crossover}.
\newblock \emph{\bibinfo{journal}{Nat. Commun.}} \textbf{\bibinfo{volume}{9}},
  \bibinfo{pages}{1914} (\bibinfo{year}{2018}).
  
\bibitem{So-prl18}
\bibinfo{author}{Kim, So Yeun} \emph{et~al.}
\newblock \bibinfo{title}{Charge-Spin Correlation in van der Waals Antiferromagnet \ce{NiPS3}}.
\newblock \emph{\bibinfo{journal}{Phys. Rev. Lett.}} \textbf{\bibinfo{volume}{120}},
  \bibinfo{pages}{136402} (\bibinfo{year}{2018}).

\bibitem{wu_flat_2007}
\bibinfo{author}{Wu, C.}, \bibinfo{author}{Bergman, D.},
  \bibinfo{author}{Balents, L.} \& \bibinfo{author}{Sarma, S.~D.}
\newblock \bibinfo{title}{Flat bands and wigner crystallization in the
  honeycomb optical lattice}.
\newblock \emph{\bibinfo{journal}{Phys. Rev. Lett.}}
  \textbf{\bibinfo{volume}{99}}, \bibinfo{pages}{070401}
  (\bibinfo{year}{2007}).

\bibitem{wu_p_2008}
\bibinfo{author}{Wu, C.} \& \bibinfo{author}{Sarma, S.~D.}
\newblock \bibinfo{title}{$p_{x,y}$-orbital counterpart of graphene: Cold atoms in
  the honeycomb optical lattice}.
\newblock \emph{\bibinfo{journal}{Phys. Rev. B}} \textbf{\bibinfo{volume}{77}},
  \bibinfo{pages}{235107} (\bibinfo{year}{2008}).

\bibitem{cao2018correlated}
\bibinfo{author}{Cao, Y.} \emph{et~al.}
\newblock \bibinfo{title}{Correlated insulator behaviour at half-filling in
  magic-angle graphene superlattices}.
\newblock \emph{\bibinfo{journal}{Nature}} \textbf{\bibinfo{volume}{556}},
  \bibinfo{pages}{80} (\bibinfo{year}{2018}).

\bibitem{chittari_electronic_2016}
\bibinfo{author}{Chittari, B.~L.} \emph{et~al.}
\newblock \bibinfo{title}{Electronic and magnetic properties of single-layer \ce{MPX3} metal phosphorous trichalcogenides}.
\newblock \emph{\bibinfo{journal}{Phys. Rev. B}} \textbf{\bibinfo{volume}{94}},
  \bibinfo{pages}{184428} (\bibinfo{year}{2016}).

\bibitem{hu_local_2012}
\bibinfo{author}{Hu, J.} \& \bibinfo{author}{Ding, H.}
\newblock \bibinfo{title}{Local antiferromagnetic exchange and collaborative
  fermi surface as key ingredients of high temperature superconductors}.
\newblock \emph{\bibinfo{journal}{Scientific Reports}}
  \textbf{\bibinfo{volume}{2}}, \bibinfo{pages}{381} (\bibinfo{year}{2012}).

\bibitem{hu_identifying_2016}
\bibinfo{author}{Hu, J.}
\newblock \bibinfo{title}{Identifying the genes of unconventional high
  temperature superconductors}.
\newblock \emph{\bibinfo{journal}{Science Bulletin}}
  \textbf{\bibinfo{volume}{61}}, \bibinfo{pages}{561--569}
  (\bibinfo{year}{2016}).

\bibitem{Seo2008}
\bibinfo{author}{Seo, K.}, \bibinfo{author}{Bernevig, B.~A.} \&
  \bibinfo{author}{Hu, J.}
\newblock \bibinfo{title}{Pairing symmetry in a two-orbital exchange coupling
  model of oxypnictides}.
\newblock \emph{\bibinfo{journal}{Phys. Rev. Lett.}}
  \textbf{\bibinfo{volume}{101}}, \bibinfo{pages}{206404}
  (\bibinfo{year}{2008}).

\bibitem{Mos2}
\bibinfo{author}{Leb\`egue, S.} \& \bibinfo{author}{Eriksson, O.}
\newblock \bibinfo{title}{Electronic structure of two-dimensional crystals from
  ab initio theory}.
\newblock \emph{\bibinfo{journal}{Phys. Rev. B}} \textbf{\bibinfo{volume}{79}},
  \bibinfo{pages}{115409} (\bibinfo{year}{2009}).

\bibitem{lee_ising-type_2016}
\bibinfo{author}{Lee, J.-U.} \emph{et~al.}
\newblock \bibinfo{title}{Ising-type magnetic ordering in atomically thin
  \ce{FePS3}}.
\newblock \emph{\bibinfo{journal}{Nano letters}} \textbf{\bibinfo{volume}{16}},
  \bibinfo{pages}{7433--7438} (\bibinfo{year}{2016}).

\bibitem{icsd_2007}
\bibinfo{author}{Allmann, R.} \& \bibinfo{author}{Hinek, R.}
\newblock \bibinfo{title}{The introduction of structure types into the
  inorganic crystal structure database icsd}.
\newblock \emph{\bibinfo{journal}{Acta. Crystallogr. A}} \textbf{\bibinfo{volume}{63}}, \bibinfo{pages}{412--417}
  (\bibinfo{year}{2007}).

\bibitem{ouvrard1985structural}
\bibinfo{author}{Ouvrard, G.}, \bibinfo{author}{Brec, R.} \&
  \bibinfo{author}{Rouxel, J.}
\newblock \bibinfo{title}{Structural determination of some \ce{MPS3} layered phases
  ({M= Mn, Fe, Co, Ni and Cd})}.
\newblock \emph{\bibinfo{journal}{Mat. Res. Bull.}}
  \textbf{\bibinfo{volume}{20}}, \bibinfo{pages}{1181--1189}
  (\bibinfo{year}{1985}).

\bibitem{klingen1968hexathio}
\bibinfo{author}{Klingen, W.}, \bibinfo{author}{Eulenberger, G.} \&
  \bibinfo{author}{Hahn, H.}
\newblock \bibinfo{title}{{\"U}ber hexathio-und hexaselenohypodiphosphate vom
  typ {M}$_{2}^{II}$\ce{P2X6}}.
\newblock \emph{\bibinfo{journal}{Naturwissenschaften}}
  \textbf{\bibinfo{volume}{55}}, \bibinfo{pages}{229--230}
  (\bibinfo{year}{1968}).

\bibitem{rao1992magnetic}
\bibinfo{author}{Rao, R.~R.} \& \bibinfo{author}{Raychaudhuri, A.}
\newblock \bibinfo{title}{Magnetic studies of a mixed antiferromagnetic system
  {F}e$_{1- x}${N}i$_x$\ce{PS3}}.
\newblock \emph{\bibinfo{journal}{J. Phys. Chem. Sol.}}
  \textbf{\bibinfo{volume}{53}}, \bibinfo{pages}{577--583}
  (\bibinfo{year}{1992}).

\bibitem{brec1980proprietes}
\bibinfo{author}{Brec, R.}, \bibinfo{author}{Ouvrard, G.},
  \bibinfo{author}{Louisy, A.} \& \bibinfo{author}{Rouxel, J.}
\newblock \bibinfo{title}{Proprietes structurales de phases {M (II)PX$_3$} ({X= S, Se}   )}.
\newblock In \emph{\bibinfo{booktitle}{Annales de chimie--science des
  mat{\'e}riaux}}, \bibinfo{pages}{499--512} (\bibinfo{year}{1980}).

\bibitem{kresse1996}
\bibinfo{author}{Kresse, G.} \& \bibinfo{author}{Furthm{\"u}ller, J.}
\newblock \bibinfo{title}{Efficiency of ab-initio total energy calculations for
  metals and semiconductors using a plane-wave basis set}.
\newblock \emph{\bibinfo{journal}{Comp. Mat. Sci.}}
  \textbf{\bibinfo{volume}{6}}, \bibinfo{pages}{15--50} (\bibinfo{year}{1996}).

\bibitem{Joubert1999FromMethod}
\bibinfo{author}{Kresse, G.} \& \bibinfo{author}{Joubert, D.}
\newblock \bibinfo{title}{From ultrasoft pseudopotentials to the projector
  augmented-wave method}.
\newblock \emph{\bibinfo{journal}{Phys. Rev. B}} \textbf{\bibinfo{volume}{59}},
  \bibinfo{pages}{1758} (\bibinfo{year}{1999}).

\bibitem{perdew_generalized_1996}
\bibinfo{author}{Perdew, J.~P.}, \bibinfo{author}{Burke, K.} \&
  \bibinfo{author}{Ernzerhof, M.}
\newblock \bibinfo{title}{Generalized gradient approximation made simple}.
\newblock \emph{\bibinfo{journal}{Phys. Rev. Lett.}}
  \textbf{\bibinfo{volume}{77}}, \bibinfo{pages}{3865--3868}
  (\bibinfo{year}{1996}).

\bibitem{mostofi2008wannier90}
\bibinfo{author}{Mostofi, A.~A.} \emph{et~al.}
\newblock \bibinfo{title}{wannier90: A tool for obtaining maximally-localised
  wannier functions}.
\newblock \emph{\bibinfo{journal}{Comp. Phys. Commun.}}
  \textbf{\bibinfo{volume}{178}}, \bibinfo{pages}{685--699}
  (\bibinfo{year}{2008}).

\bibitem{Dudarev1998Electron-energy-lossStudy}
\bibinfo{author}{Dudarev, S.~L.}, \bibinfo{author}{Botton, G.~A.},
  \bibinfo{author}{Savrasov, S.~Y.}, \bibinfo{author}{Humphreys, C.~J.} \&
  \bibinfo{author}{Sutton, A.~P.}
\newblock \bibinfo{title}{Electron-energy-loss spectra and the structural
  stability of nickel oxide: An {LSDA+U} study}.
\newblock \emph{\bibinfo{journal}{Phys. Rev. B}} \textbf{\bibinfo{volume}{57}},
  \bibinfo{pages}{1505} (\bibinfo{year}{1998}).

\bibitem{Liyinxiang2017}
\bibinfo{author}{Li, Y.} \emph{et~al.}
\newblock \bibinfo{title}{Robust $d$-wave pairing symmetry in multiorbital
  cobalt high-temperature superconductors}.
\newblock \emph{\bibinfo{journal}{Phys. Rev. B}} \textbf{\bibinfo{volume}{96}},
  \bibinfo{pages}{024506} (\bibinfo{year}{2017}).

\bibitem{Scalapino2012}
\bibinfo{author}{Scalapino, D.~J.}
\newblock \bibinfo{title}{A common thread: The pairing interaction for
  unconventional superconductors}.
\newblock \emph{\bibinfo{journal}{Rev. Mod. Phys.}}
  \textbf{\bibinfo{volume}{84}}, \bibinfo{pages}{1383--1417}
  (\bibinfo{year}{2012}).

\bibitem{wildes_magnetic_2015}
\bibinfo{author}{Wildes, A.} \emph{et~al.}
\newblock \bibinfo{title}{Magnetic structure of the quasi-two-dimensional
  antiferromagnet nips 3}.
\newblock \emph{\bibinfo{journal}{Phys. Rev. B}} \textbf{\bibinfo{volume}{92}},
  \bibinfo{pages}{224408} (\bibinfo{year}{2015}).

\bibitem{castellani-prb78}
\bibinfo{author}{Castellani, C.}, \bibinfo{author}{Natoli, C.~R.} \&
  \bibinfo{author}{Ranninger, J.}
\newblock \bibinfo{title}{Magnetic structure of \ce{V2O3} in the insulating phase}.
\newblock \emph{\bibinfo{journal}{Phys. Rev. B}} \textbf{\bibinfo{volume}{18}},
  \bibinfo{pages}{4945--4966} (\bibinfo{year}{1978}).

\bibitem{kk}
\bibinfo{author}{Kugel, K.} \&
  \bibinfo{author}{Khomskii, D.}
\newblock \bibinfo{title}{The Jahn-Teller effect and magnetism: transition metal compounds}.
\newblock \emph{\bibinfo{journal}{Physics-Uspekhi}}
  \textbf{\bibinfo{volume}{25}}, \bibinfo{pages}{231} (\bibinfo{year}{1982}).

\bibitem{QHWang-prb04}
\bibinfo{author}{Wang, Q.-H.}, \bibinfo{author}{Lee, D.-H.} \&
  \bibinfo{author}{Lee, P.~A.}
\newblock \bibinfo{title}{Doped $t\ensuremath{-}J$ model on a triangular
  lattice: Possible application to
  ${\mathrm{Na}}_{x}{\mathrm{CoO}}_{2}\ensuremath{\cdot}y{\mathrm{H}}_{2}\mathrm{O}$
  and ${\mathrm{Na}}_{1\ensuremath{-}x}{\mathrm{TiO}}_{2}$}.
\newblock \emph{\bibinfo{journal}{Phys. Rev. B}} \textbf{\bibinfo{volume}{69}},
  \bibinfo{pages}{092504} (\bibinfo{year}{2004}).

\bibitem{PatrickLee01renormalized}
\bibinfo{author}{Brinckmann, J.} \& \bibinfo{author}{Lee, P.~A.}
\newblock \bibinfo{title}{Renormalized mean-field theory of neutron scattering
  in cuprate superconductors}.
\newblock \emph{\bibinfo{journal}{Phys. Rev. B}} \textbf{\bibinfo{volume}{65}},
  \bibinfo{pages}{014502} (\bibinfo{year}{2001}).

\bibitem{Sigrist05slave}
\bibinfo{author}{R{\"u}egg, A.}, \bibinfo{author}{Indergand, M.},
  \bibinfo{author}{Pilgram, S.} \& \bibinfo{author}{Sigrist, M.}
\newblock \bibinfo{title}{Slave-boson mean-field theory of the mott transition
  in the two-band hubbard model}.
\newblock \emph{\bibinfo{journal}{Eur. Phys. J. B}} \textbf{\bibinfo{volume}{48}},
  \bibinfo{pages}{55--64} (\bibinfo{year}{2005}).

\bibitem{li2018new}
\bibinfo{author}{Li, W.} \emph{et~al.}
\newblock \bibinfo{title}{A new superconductor of cuprates with unique
  features}.
\newblock \emph{\bibinfo{journal}{arXiv:1808.09425}}  (\bibinfo{year}{2018}).

\bibitem{maier2018d}
\bibinfo{author}{Maier, T.~A.}, \bibinfo{author}{Berlijn, T.} \&
  \bibinfo{author}{Scalapino, D.~J.}
\newblock \bibinfo{title}{$ d $-wave and $ s_{\pm}$ pairing strengths in \ce{Ba2CuO}$_{3+\delta}$}.
\newblock \emph{\bibinfo{journal}{arXiv:1809.04156}}  (\bibinfo{year}{2018}).

\bibitem{jiang2018nodeless}
\bibinfo{author}{Jiang, K.}, \bibinfo{author}{Wu, X.}, \bibinfo{author}{Hu, J.}
  \& \bibinfo{author}{Wang, Z.}
\newblock \bibinfo{title}{Nodeless high-{T}$_c $ superconductivity in
  highly-overdoped monolayer \ce{CuO2}}.
\newblock \emph{\bibinfo{journal}{arXiv:1804.05072}}  (\bibinfo{year}{2018}).

\bibitem{le_possible_2017}
\bibinfo{author}{Le, C.}, \bibinfo{author}{Zeng, J.}, \bibinfo{author}{Gu, Y.},
  \bibinfo{author}{Cao, G.-H.} \& \bibinfo{author}{Hu, J.}
\newblock \bibinfo{title}{A possible family of {Ni}-based high temperature
  superconductors}.
\newblock \emph{\bibinfo{journal}{Science Bulletin}}
  \textbf{\bibinfo{volume}{63}}, \bibinfo{pages}{957 -- 963}
  (\bibinfo{year}{2018}).

\bibitem{Ye2009}
\bibinfo{author}{Ye, J.~T.} \emph{et~al.}
\newblock \bibinfo{title}{Liquid-gated interface superconductivity on an
  atomically flat film}.
\newblock \emph{\bibinfo{journal}{Nat. Mater.}}
  \textbf{\bibinfo{volume}{9}}, \bibinfo{pages}{125} (\bibinfo{year}{2009}).

\end{thebibliography}

\end{document}